\documentclass[twocolumn]{pasj00}
\renewcommand{\tabcolsep}{4pt}

\usepackage{lscape}

\begin{document}

\title{ Near Infrared Spectroscopy of M Dwarfs. II.\\
H$_2$O Molecule as an Abundance Indicator of Oxygen\thanks{Based on
data collected at Subaru Telescope, which is operated by the National
Astronomical Observatory of Japan.} }

\author{Takashi \textsc{Tsuji}} %
\affil{Institute of Astronomy, School of Science, The University of Tokyo,
2-21-1 Osawa, Mitaka-shi, Tokyo, 181-0015}
\email{ttsuji@ioa.s.u-tokyo.ac.jp }

\author{Tadashi \textsc{Nakajima}}
\affil{National Astronomical Observatory of Japan, 2-21-1 Osawa,
Mitaka-shi, Tokyo, 181-8588}
\email{tadashi.nakajima@nao.ac.jp}

\and

\author{Yoichi \textsc{Takeda}}
\affil{National Astronomical Observatory of Japan, 2-21-1 Osawa,
Mitaka-shi, Tokyo, 181-8588}
\email{takeda.yoichi@nao.ac.jp}


\maketitle

\begin{abstract}
Based  on the near infrared  spectra ($R \approx 20000$) of M dwarfs,
oxygen abundances are determined from the ro-vibrational lines 
of H$_2$O. Although H$_2$O lines in M dwarfs are badly blended each other 
and the continuum levels are depressed appreciably by the collective 
effect of numerous  H$_2$O lines themselves, quantitative analysis 
of H$_2$O lines has been carried out by referring to the pseudo-continua 
consistently defined by the same way on the observed and theoretical spectra.
For this purpose, the pseudo-continuum on the theoretical spectrum has
been evaluated accurately by the use of the recent high-precision H$_2$O
line-list. Then, we propose a simple and flexible method of analyzing 
equivalent widths (EWs) of blended features (i.e., not necessarily limited 
to single lines) by the use of a mini curve-of-growth (CG), which is a small 
portion of the usual CG around the observed
EW. The mini CG is generated by using the theoretical  EWs 
evaluated from the synthetic spectrum by exactly the same way as the EWs
are measured from the observed spectrum. The observed EW is
converted to the abundance by the use of the mini CG, and the process 
is repeated for all the observed EWs line-by-line or blend-by-blend. 

In cool M dwarfs, almost all the oxygen atoms left after CO formation
are in stable H$_2$O molecules, which suffer little change for the
uncertainties due to imperfect  modelling of the  photospheres. 
Moreover, the thermal
velocity of H$_2$O is most probably larger than the micro-turbulent
velocity because of its lower molecular weight, and the uncertainty
of the micro-turbulent  velocity will have relatively minor effect on the
abundance determination. Then the numerous H$_2$O lines
are excellent abundance indicators of oxygen. 
The oxygen abundances are determined to be 
log\,$A_{\rm O}$  ($A_{\rm O} = N_{\rm O}/N_{\rm H}$) between -3.5 and 
-3.0 in 38 M dwarfs, but cannot 
in four early M dwarfs in which H$_2$O lines are detected only marginally. 
The resulting log\,$A_{\rm O}/A_{\rm C}$  plotted against 
log\,$A_{\rm C}$ appears to be systematically smaller in the carbon-rich 
M dwarfs,  showing the different formation histories of oxygen 
and carbon in the chemical evolution of the Galactic disk. 
Also, $A_{\rm O}$/$A_{\rm Fe}$ ratios in most M dwarfs are closer to the 
solar $A_{\rm O}$/$A_{\rm Fe}$ ratio
based on the classical high oxygen abundance rather than on the recently 
downward revised low value.
 
\end{abstract}

\section{Introduction}

Presence of water in stellar photospheres has been predicted based on 
the theory of thermochemistry (e.g. \cite{Rus34}),
and the observational confirmation has been done on a Mira variable star at 
the infancy of the infrared spectroscopy \citep{Kui62}.    
However, detailed observations of water vapor in celestial objects
have been hampered by the obscuration of the water vapor in the
Earth's atmosphere. For this reason, the most clear demonstration
of the water vapor in stellar atmospheres has been done by the balloon-born 
telescope known as Stratoscope II on several red giant and supergiant
stars \citep{Woo64}. This pioneering undertaking had been taken over  
after 30 years by the Infrared Space 
Observatory ISO \citep{Kes96}, which revealed that water exists 
everywhere in the Universe. Now, water (H$_2$O) may be assumed to be the third 
abundant molecule next to H$_2$ and CO in the Universe.  

Meanwhile, important progress on the high resolution infrared spectroscopy
has been achieved already in the 1970's, despite the rather noisy 
infrared detectors at that time.  This has been made possible
with the Fourier Transform Spectroscopy (FTS) pioneered by  
\citet{Con70} and further developed at Kitt Peak National 
Observatory (e.g., \cite{Rid84}).  
Based on observation of cool M giant  Mira  variable stars with the FTS
of KPNO 4m Telescope, 
a detailed study on the high resolution H$_2$O spectra on celestial object
outside the solar system has been done by \citet{Hin79} for the first time,
with the use of the molecular data on  H$_2$O  known at that time.
Their analysis based on the curve-of-growth method revealed complicated 
behaviors of H$_2$O spectra, and showed the presence of a very cool 
layer different from the pulsating photospheric layer. 
This first attempt at analyzing the spectra of H$_2$O
in M-type Miras  showed that the H$_2$O spectra contain a wealth of 
information on the coolest outer atmosphere where stellar photosphere
and cool circumstellar layers interact. 

Next detailed study on high resolution celestial H$_2$O spectra has been
initiated with the detection of many H$_2$O lines on sunspot umbral
spectra with the FTS of the National Solar Observatory at Kitt Peak
(\cite{Wal92}; \cite{Wal95}), followed by detailed laboratory 
analyses of hot H$_2$O spectra (\cite{Pol97}; \cite{Zob00}; \cite{Ter02}). 
Then detailed laboratory and  theoretical works on hot water 
have been continued by many groups as reviewed by \citet{Ber02}.  As 
the fruits, an extended line-list of hot water based on the laboratory 
analysis has been completed \citep{Zob08}.
Also,  a high-accuracy H$_2$O line-list comprising over 500 million 
transitions has been completed based on an {\it ab initio} approach 
\citep{Bar06}, and compiled in large databases such as 
HITEMP2010 \citep{Rot10}. 
The recent progress achieved in the molecular database constitutes 
the most important basis for the high resolution molecular spectroscopy 
of celestial objects.

On the other hand,  H$_2$O molecule has been recognized  as an important 
source of opacity in  cool gaseous mixture such as the photospheres of 
M-type stars (e.g., \cite{Tsu66}; \cite{Aum67}), and the H$_2$O opacity 
has been applied to 
initial attempts of constructing model photospheres of M dwarfs 
(e.g., \cite{Aum69}; \cite{Tsu69}).  The observational manifestations
of water in M dwarfs have been done from low 
resolution spectra  and  progressive importance of H$_2$O as an infrared
opacity source in cooler M dwarfs has been well established 
(e.g., \cite{Ber87}; \cite{Tin93}; \cite{Jon94}). 

 The high resolution spectroscopy has been more difficult for such faint 
objects as M dwarfs until recently. Even with the FTS, M dwarfs were too 
faint and we know only one pioneering attempt at analyzing the infrared 
high resolution spectra of M dwarfs by \citet{Mou78}. 
However, recent progress in infrared spectroscopy with the new infrared 
detectors finally made it possible to observe infrared spectra of 
faint objects including M dwarfs at high resolution (e.g., \cite{One12}).
 We started to explore such  a new possibility 
opened by the progress of infrared spectroscopy with the use of
the echelle mode of InfraRed Camera and Spectrograph, IRCS 
\citep{Kob00} at Subaru. 

Although the fundamental parameters such as the effective temperatures
needed for  spectroscopic analysis were poorly known for M dwarfs 
until recently, recent progress of stellar interferometry
has finally resolved this problem and accurate effective temperatures
are now known based on the measured angular diameters.  
We found that  the effective temperatures, $T_{\rm eff}$, based on the 
interferometry (\cite{Boy12} and references cited therein) and 
supplemented by those based on the infrared flux method for cooler M 
dwarfs \citep{Tsu96}, show a fairly tight 
correlation with the absolute magnitudes at 3.4 $\mu$m, $M_{3.4}$, based 
on the WISE data \citep{Wri10}. Then we proposed a simple method to  
infer $T_{\rm eff}$ from $M_{3.4}$ available to almost all the M dwarfs 
we have observed (\cite{Tsu14}; hereafter be referred to as Paper I).

Another problem in the quantitative analysis of the spectra of M dwarfs
is that  the continuum levels cannot be seen  because of the veil 
opacity due to numerous molecular lines, especially of H$_2$O. 
For this difficulty, we showed that the spectroscopic analysis of M dwarfs
can be done by referring to the pseudo-continuum levels both on the 
observed and theoretical spectra. This is possible since the pseudo-continuum
level on the theoretical spectrum can be evaluated accurately thanks to the 
recently developed  high precision H$_2$O line database (Paper I). 
It can be  shown   that the quantitative 
analysis of the spectrum referring to the pseudo-continuum is 
essentially the same as that referring to the true-continuum. 

 In this paper, we use the observed data introduced in Paper I (as for
detail, see its Table 1) in which we analyzed the CO spectra.
In this paper, we focus our attention to the H$_2$O spectra. We first
summarize the basic physical parameters and model photospheres of M dwarfs
we study in this paper (section 2). Then,  
we examine the H$_2$O spectra in the $K$ band region (section 3). We 
analyze a portion of the spectral region with the strong H$_2$O features 
in the midst of the 1.9 $\mu$m band (section 4) and another portion with
relatively sparse H$_2$O blends in the
tail of the H$_2$O 2.7 $\mu$m band (section 5). 
We will show that H$_2$O is an excellent abundance indicator of  oxygen 
 in M dwarfs (subsection 6.1) and  oxygen abundances in 38 M dwarfs have
been determined (subsection 6.2). We summarize the method of
abundance analysis which is essentially very simple (subsection 6.3). We 
close our discussion with a prospect on the spectroscopy of H$_2$O in 
stellar spectra other than M dwarfs (subsection 6.4).

\section{Fundamental Parameters and Model Photospheres of M Dwarfs}

We summarize the basic parameters of our 42 program stars in Table\,1.
Following the object names in the first column, the spectral types
given in the second column are from \citet{Joy74} for 29 M dwarfs 
(those beginning with dM in Table\,1), from SIMBAD for the  
12 cases (those beginning with M), and  unknown for the remaining 
one object (GJ\,768.1C). The values of $T_{\rm eff}$ and log\,$g$ given
in the third and fourth columns, respectively, are those  used in
our analysis of CO in Paper I. In discussing the values of $T_{\rm eff}$,
however, we used the WISE All-Sky Release in Paper I to obtain 
 $M_{3.4}$, the absolute magnitude based on the WISE $W1$ flux,
throughout. Later we notice that there is a new version referred to as
the AllWISE Catalog. According to the AllWISE Explanatory 
Supplement\footnote
{http://wise2.ipac.caltech.edu/docs/release/allwise/expsup.},  
the $W1$ and $W2$ photometric sensitivity is better in the AllWISE Catalog
than in the WISE All-Sky data we have applied in Paper I.
However, it is also noted that the sources brighter than $W1 < 8$\,mag
may have better photometry in the WISE All-Sky Catalog
than in the AllWISE Catalog. 

We have examined the  AllWISE data and find that the results 
for some K and M dwarfs we have used as calibrators of  
$M_{3.4} - {\rm log}\,T_{\rm eff}$ relation in Paper I were changed to 
unreasonable values that may suffer the saturation effect or even some 
objects disappeared from the database. As a result, the number of 
calibration sources using $T_{\rm eff}$ values by interferometry 
\citep{Boy12} decreases from 27 to 15 and those using $T_{\rm eff}$ values
by the infrared flux method \citep{Tsu96} from 9 to 7.
But we add three new M dwarfs for which angular diameters are measured 
anew \citep{Bra14} and thus the number of calibration stars is 
15 + 7 + 3 = 25 as shown in Tables\,10 \& 11 (Appendix I). These data of 
Tables\,10 \& 11 are plotted in Fig.\,18 (Appendix I), on which the mean  
$M_{3.4} - {\rm log}\,T_{\rm eff}$ relation from Fig.\,1 in Paper I
is reproduced.  Despite some changes in the WISE database, the AllWISE data
are consistent with the mean $M_{3.4} - {\rm log}\,T_{\rm eff}$ relation based
on the WISE All-Sky Catalog. We reanalyze a few objects with $F_{3.4} > 8$ 
mag in Table\,5 of Paper I with the data of the AllWISE, but the resulting 
$T_{\rm eff}$ values agree within 10\,K with the results based on the WISE 
All-Sky data. Since most of M dwarfs in Table 5 of Paper I are brighter 
than $F_{3.4} = 8$ mag, we decide to use the results of Table\,5 in Paper I 
to all the objects\footnote{This also applies to the values of $T_{\rm eff}$ 
of three objects for which angular diameters have been measured \citep{Bra14}
after Paper I was completed, for consistency of our analysis with that of 
Paper I. Also, we wonder why $T_{\rm eff}$ of GJ\,876 by the new 
interferometric measurement deviates so large from the mean 
$M_{3.4} - {\rm log}\,T_{\rm eff}$ relation, as shown in Fig.\,18
 (Appendix I; also see Table\,14 in Paper I), as the case of GJ\,725B 
(see Fig.\,1 in Paper I).}. 

We also apply the same model photospheres used in the analysis of CO in 
Paper I, and these models are given in the fifth column of Table\,1. The 
model photospheres are designated by cloud type/abundance case/$T_{\rm eff}$
/log\,$g$ as in our model database referred to as the unified cloudy model
(UCM)\footnote{http://www.mtk.ioa.s.u-tokyo.ac.jp/$\sim$ttsuji/export/ucm2.}.  
The cloud type is defined to characterize the
thickness of the dust cloud formed in the photospheres of cool dwarfs.
The dust cloud forms at the condensation temperature $T_{\rm cond}$
and dissolves at a critical temperature $T_{\rm cr}$ because dust particles 
become too large and precipitate. Thus dust cloud  exists in the region
of $ T_{\rm cr} \lesssim T \lesssim T_{\rm cond}$, where $T_{\rm cond}$ is 
fixed  thermochemically. Then, thickness of the cloud depends on $T_{\rm cr}$
which is a free parameter in our UCMs and used to define the cloud type    
(as for details see \cite{Tsu02}; \yearcite{Tsu05}).
The cloud type includes a limiting case of no dust cloud (referred to 
as a clear case, C). Generally, dust clouds are formed in the photospheres of 
cool dwarfs with  $T_{\rm eff}$ below about 2600\,K, and hence all our 
present sample are dust-free. Also, {\it case a} abundance  is based 
on the classical solar abundance with log\,$A_{\rm C}$ = -3.40\footnote
{We use the notation: $A_{\rm El} =
N_{\rm El}/N_{\rm H}$, where $N_{\rm El}$ and $N_{\rm H}$ are the number 
densities of the element El and hydrogen, respectively.} 
and log\,$A_{\rm O}$ = -3.08 (see Table\,1 in \cite{Tsu02}), and {\it case c}
on the downward revised solar C \& O abundances of log\,$A_{\rm C}$ = -3.61
and log\,$A_{\rm O}$ = -3.31 \citep{All02}.  We used the UCM grid in
our preliminary analysis, but generated specified model for $T_{\rm eff}$
and log\,$g$ of each object in the subsequent analysis. These models  
are referred to, for example,  as Ca3570c489 implying a clear model
without dust cloud, with the {\it case a} abundance, $T_{\rm eff}$ =
3570\,K, and log\,$g$ = 4.89.

\vspace{2mm}
--------------------------------------------------------------------

Table 1: Fundamental parameters and model photospheres (p.27).

--------------------------------------------------------------------

\section{H$_2$O in the Spectra of M Dwarfs}

\subsection{Molecular Data of H$_2$O}
We first try to identify H$_2$O  transitions by referring to the 
laboratory data by \citet{Zob08}, but we soon notice that the 
BT2-HITEMP2010 database (\cite{Bar06}; \cite{Rot10}), which we have already 
examined in Paper I, provides fairly accurate line positions of H$_2$O 
lines. It has generally been thought that the {\it ab initio}
approach to the molecular structure and spectroscopic data provides 
large line-lists but their  accuracy especially on the line
positions cannot be very high. However, the recently computed line-list by
\citet{Bar06} attained an accuracy to be used for identifications of H$_2$O 
lines at last. Since intensity data are also required for abundance 
analysis, we decide to use the BT2-HITEMP2010 data in the following 
analysis. The accurate laboratory data should certainly be useful 
if we are to study radial velocities, for example. 

Under the high density of the photospheres of M dwarfs, pressure
broadening plays an important role.
Usually, collision half-width $ \gamma$ is represented by    
   $$ \gamma = \gamma_{0}{p \over p_{0} }( {T_{0} \over T}  )^{n} ,
\eqno (1)    $$
where $\gamma_{0}$ is the collision half-width measured at a reference
temperature $T_{0}$ (e.g. 296\,K) and gas pressure $p_{0}$.
As to pressure broadening coefficients for H$_2$O perturbed by H$_2$ 
and He relevant to stellar photospheres, there are some experimental 
(e.g., \cite{Ste04}, \cite{Fau13}) and theoretical (e.g., \cite{Gam96})
works for pure rotational transitions. 
We have compared the resulting collision half-widths by the H$_2$ 
and He broadening with those by the air broadening for the case of CO
and found that they are not drastically different in Table\,6 of Paper I.  
The case of H$_2$O is found to be more or less the same,
and we assume a median value of 
$\gamma_{0}$ = 0.08\,cm$^{-1}$\,atm$^{-1}$ by the air broadening 
(e.g., \cite{Rot10}) for all the H$_2$O lines in M dwarfs.

\subsection{H$_2$O Spectra in the K Band Region}
Water molecule is a tri-atomic asymmetric top molecule having three normal 
vibration modes
with $\nu_{1}$ = 3651.7\,cm$^{-1}$, $\nu_{2}$ = 1595.0\,cm$^{-1}$, and
$\nu_{3}$ = 3755.8\,cm$^{-1}$ \citep{Her45}. In the $K$ band region, 
the H$_2$O 2.7 $\mu$m band is mainly composed of the  
$\nu_{1}$ and $\nu_{3}$ fundamentals and 2$\nu_{2}$ overtone, while
the H$_2$O 1.9\,$\mu$m band  mainly of the combination bands 
$\nu_{2} + \nu_{3}$ and $\nu_{1} + \nu_{2}$. The  model spectra of H$_2$O
(with resolution of $R \approx 2000$) based on the BT2-HITEMP2010 line-list
for four M dwarf model photospheres of $T_{\rm eff}$ = 2800, 3200, 3600, 
and 4000\,K, assuming the solar metallicity ({\it case a} ), are shown 
in Fig.\,1. The 1.9\,$\mu$m band is still very weak in the model of 
$T_{\rm eff}$ = 4000\,K and strengthens towards cooler models.
The 2.7\,$\mu$m band is already visible weakly in the model of 
$T_{\rm eff}$ = 4000\,K and appears to be very strong in cooler models. 

We choose two portions noted as region A (20296--20391\,{\AA}) and
region B (22515--22935\,{\AA}) in Fig.\,1 for our analysis of H$_2$O
spectra.  At the top of Fig.\,1, the atmospheric window (the $K$ band
window) in which atmospheric transmission is larger than 50\,\% under
conditions appropriate for Mauna Kea, Hawaii, is indicated based on
the data given by \citet{Cox99}.  Both the regions A and B are in the
atmospheric window.  The region A is in the midst of the 1.9\,$\mu$m
band where H$_2$O lines are strong. We select this particular region
because the atmospheric absorption appears to be relatively weak by
visual inspection of the observed spectra. The region B is disturbed
by both the 1.9 and 2.7\,$\mu$m bands least, but nevertheless some
H$_2$O features are found in this region in later M dwarfs. Since
H$_2$O lines in the region A are very strong in later M dwarfs and the
pseudo-continua are depressed by as much as 15\,\% from the true
continua\footnote{It is of course not possible to see the
  true-continuum level in the observed spectrum of M dwarf, but it can
  be estimated with the help of the theoretical spectrum, as will be
  shown in subsection\,4.3.}, we think it's useful to analyze the
H$_2$O lines without such heavy blendings for comparison. The H$_2$O
blends are sparsely distributed in the region B and we survey a larger
spectral range to measure a modest number of H$_2$O blends.

By the way, we have confined our analysis of CO lines in Paper I
to the bandhead region of the CO 2-0 band, and
 it is confirmed in Fig.\,1 that this region adjacent to the region B 
is disturbed by both the H$_2$O 1.9 and 2.7\,$\mu$m bands least and
that it may be more difficult to analyze CO spectra in other regions,
because of the severe blending of H$_2$O lines.

\vspace{2mm}
--------------------------------------------------------------------

Fig.\,1: Model spectra of M dwarfs in the $K$ band region (p.14).

--------------------------------------------------------------------

\subsection{H$_2$O Spectra in the Region A}
As examples of the observed spectra in the region A near 2.0 $\mu$m,
the cases of four M dwarfs are shown in Fig.\,2.
We select candidates of  H$_2$O blends to be measured for abundance
analysis and indicate them by arrows with reference numbers A01--A17.
 The total number of H$_2$O lines in the region A (20296--20391\,{\AA})
from BT2-HITEMP2010 line-list with the cut off at the integrated intensity
$S(T=3000\,{\rm K}) \approx 3.10^{-27}$cm$^{-1}$\,molecule$^{-1}$ is 69909 
lines  or about 736\,lines per 1\,{\AA} interval, and
we only show the strong line(s) near the position indicated by each
arrow in Table\,2\footnote{It must be noted that this line-list can
never be used for the actual analysis of the spectra and for generating the
synthetic spectra, since the effect of the underlying and nearby weak H$_2$O 
lines are quite appreciable. The full line-list including all 
the weak lines must be applied even to evaluate the spectrum of a single 
blend feature.}.  Following the reference number and the observed
wavelength (in vacuum) in the columns 1 and 2, respectively, the wavelength,
assignments of the upper and lower vibrational levels  and of the upper and
lower rotational levels, log\,$gf$ value, and the lower excitation
potential (in cm$^{-1}$)  from the BT2-HITEMP2010 are given successively
after the column 3. The vibrational level is defined by the three quantum
numbers $v_{1}v_{2}v_{3}$  representing the vibrational states of the three
fundamental vibrations. The rotational level is defined by the three
parameters $J, K_{a}$, and $K_{c}$: $J$ is the rotational quantum
number of the total angular momentum, $K_a$ and $K_c$ are the quasi-quantum
numbers that in the limiting cases of the prolate and oblate symmetric 
rotors would become the the quantum numbers of the angular momenta 
associated with the respective figure axes \citep{Kin43}. 

In the early M dwarf GJ\,338A shown in Fig.\,2a, H$_2$O lines are still weak
as can be inferred from Fig.\,1, but almost all the  features shown by the 
arrows are definitely confirmed as due to H$_2$O lines  in this dM0.5 dwarf.
However, in other early type M dwarfs, GJ\,380 (dM0.5), GJ\,820B (dM0), 
GJ\,884 (dM0.5), and HIP\,12961 (M0), these H$_2$O features are
not necessarily very clear and cannot definitely be identified,
even though some of the H$_2$O features can be recognized.
Except for these four M dwarfs, the features indicated by the arrows are 
definitely confirmed as due to H$_2$O lines
in 38 M dwarfs out of our 42 M dwarfs. For examples, H$_2$O features are well 
developed in the dM2.5 dwarf GJ\,15A, shown in Fig.\,2b, and increasingly
stronger in later M dwarfs as in the dM4 dwarf GJ\,611B shown in Fig.\,2c.
In the coolest M dwarf GJ\,406 in our sample, H$_2$O features are
very strong as shown in Fig.\,2d. 
The spectrum of the B type dwarf Regulus ($\alpha$ Leo, B7V) shown in     
Fig.\,2e reveals the telluric lines mostly due to  water vapor.
Even though the region A is in the midst of  the H$_2$O 1.9\,$\mu$m band,
the telluric H$_2$O lines are not so strong in this region A within the
atmospheric K window, and have been removed in the observed
spectra of M dwarfs rather well during the 
data reduction with Regulus as calibrator.  

Finally, we measure the equivalent widths (EWs) of the H$_2$O blends 
by referring to the pseudo-continuum of each object. We select those
features whose profiles  are well defined both on their left and right
wings and, as a result, not all the
blends marked by arrows in Fig.\,2 are measured. The resulting values of 
log\,$(W/\lambda)_{\rm obs}$ (where $W$ is equivalent width)  for the H$_2$O
blends with the reference numbers indicated are given 
in Table 3. 

\vspace{2mm}
--------------------------------------------------------------------

Fig.\,2: H$_2$O in the region A of the observed spectra (p.15).

\vspace{2mm}

Table\,2: H$_2$O blends in the region A (p.28).

Table\,3: log\,$(W/\lambda)_{\rm obs}$ of H$_2$O blends in the region A
          (for online version only) (p.29).

---------------------------------------------------------------------

\subsection{H$_2$O Spectra in the Region B}
As an example of the observed spectrum in the region B near 2.3 $\mu$m,
the case of the coolest M dwarf in our sample, GJ\,406 (dM6.5), is shown in 
Fig.\,3. In this region, we may not expect to find many stellar H$_2$O lines 
from Fig.\,1. However, we already know that there are many weak H$_2$O
features in the region of CO 2-0 band just adjacent to the region B (Paper I) 
and we find that there are considerable number of weak features possibly 
due to H$_2$O blends in our higher resolution spectra ($R \approx 20000$ 
compared to $R \approx 2000$ of Fig.\,1) of later M dwarfs. We mark these 
features again by arrows in Fig.\,3 with reference numbers B01--B27. 
The features marked by arrows are in fact mostly identified as due 
to stellar H$_2$O lines by referring to the BT2-HITEMP2010 line-list. 
The transition(s) that give the large contribution(s) to each feature are 
given in Table\,4, in the same format as in Table\,2.
The total number of H$_2$O lines in the region B (22515 -- 22935\,{\AA})
from BT2-HITEMP2010 line-list with the cut-off at $S(T=3000\,{\rm K})
 \approx 3.10^{-27}$\,cm$^{-1}$\,molecule$^{-1}$ is 317621\,lines or 
about 756\,lines per 1\,{\AA} interval.
Thus it is not possible to measure the isolated single line of H$_2$O
even in this region where blendings are expected to be not so heavy
compared to the region A\footnote{But the line density in the region B
is nearly the same or even larger compared to that in the region A. This is 
because there are many weak lines even in the region where strong H$_2$O 
lines are scarce as in the region B.}, and we had to analyze the blended 
H$_2$O features throughout this paper.

In the spectrum of the region B shown in Fig.\,3, some strong features
are known as due to atomic lines of Ca I in the sunspot umbral spectrum 
\citep{Wal92}, and these features are also noted on Fig.\,3. The spectrum 
of Regulus is also shown by the solid (green or grey) line 
in Fig.\,3. There are some strong telluric lines in the region B,
and they are due to telluric CH$_4$ and N$_2$O rather than H$_2$O
\citep{Moh55}. The effect of these telluric absorption has again been 
removed by Regulus as calibrator during the data reduction, and such a 
removal of telluric absorption has been done fairly well in general
possibly because the telluric absorption in this region is not so
strong. In fact, the removal of the telluric absorption in the regions 
with strong telluric absorption (of over 90\% absorption) cannot be
done well and some residuals remain occasionally, while inspection 
of the corrected spectra in the regions A and B does not show such residuals.
   
We also measure the equivalent widths (EWs) of the H$_2$O blends 
by referring to the pseudo-continuum of each object. 
We can measure the weak H$_2$O blends in the region B in 20 out of 
42 M dwarfs in our sample. These M dwarfs are mostly later than dM3.5 
except for the dM2.5 dwarfs pair GJ\,797B or 
are M dwarfs with $T_{\rm eff}$ lower than 3500\,K.   
The resulting values of log\,$(W/\lambda)_{\rm obs}$ for the H$_2$O
blends with the reference numbers indicated are given in Table 5. 

\vspace{2mm}
--------------------------------------------------------------------

Fig.\,3: H$_2$O in the region B of the observed spectrum of GJ\,406 (p.16).

\vspace{2mm}

Table\,4: H$_2$O blends in the region B (p.30).

Table\,5: log\,$(W/\lambda)_{\rm obs}$ of H$_2$O blends in the region B
          (for online version only) (p.31).

---------------------------------------------------------------------

\section{Analysis of  H$_2$O Lines in the Region A }

\subsection{Mini Curves-of-Growth Analysis of H$_2$O Blends}

Of the four parameters that characterize stellar spectra, the effective
temperature and gravity are known (see Table\,1 based on Tables 3, 4, \& 5
of Paper I), and we assume the micro-turbulent velocity to be 
$\xi_{\rm micro}$ = 1\,km\,sec$^{-1}$
as in Paper I. Then, the only unknown parameter to be determined from the
spectra is abundance. Given that the carbon abundances are already known 
in Paper I (reproduced in Table\,6), rough estimation of the oxygen 
abundances can be possible.  
For example, we assume a starting oxygen abundance (logarithmic) to be
 $$ {\rm log}\,A_{\rm O}^{0} =  {\rm log}\,A_{\rm C} + 0.30,  \eqno(2) $$  
where ${\rm log}\,A_{\rm C}$ is the logarithmic carbon abundance from Paper I.
Note that this relationship is taken from the solar case \citep{All02}. 

Then, we apply a simple method using mini curve-of-growth blend-by-blend:
We generate  synthetic spectra for $ {\rm log}\,A_{\rm O}^{0} $
and $ {\rm log}\,A_{\rm O}^{0} + \delta $ with  the carbon abundance 
and the specific model photosphere for each object from Paper I
(also noted in Table\,1).
We generally assume $\delta = \pm0.1 $ at the beginning.
For the H$_2$O blends for which EWs are measured (Table\,3), we
evaluate  EWs, $W(\delta)$ ($\delta = 0, \pm 0.1 $), from the three synthetic 
spectra we have just prepared, by
exactly the same way as we measure the EWs from the observed spectrum.
Especially, we refer to the pseudo-continua generally defined consistently
both on the observed (see Fig.\,2) and predicted spectra.

Now, we have  a theoretical mini curve-of-growth  defined by
log\,$(W(\delta)/\lambda) $ vs. $\delta = -0.1, 0.0,$ and +0.1, where
the values of $W(\delta)$ are evaluated from the synthetic spectra
as noted above.　   
The value of the abundance correction $\Delta{\rm log}\,A_{\rm O}$ 
 to be consistent with  the observed value of log\,$(W/\lambda)_{\rm obs}$ 
from Table\,3 can be determined by the use of this mini curve-of-growth. 
This process is repeated blend-by-blend for all the H$_2$O blends 
measured. If the resulting values of $\Delta{\rm log}\,A_{\rm O}$
 for many blends appear to be less than -0.1, for example,
we generate a synthetic spectrum for $\delta = -0.2$ and 
reanalyze the same data by applying  mini curves-of-growth defined by
log\,$(W(\delta)/\lambda)$ vs. $\delta =  -0.2, -0.1$, and  0.0\footnote{
We generally do not apply 3-points interpolation formula, but apply
linear interpolation averting extrapolation as far as possible.}.
Finally, we derive the mean abundance correction from the values of 
$\Delta{\rm log}\,A_{\rm O}$ by all the measured blends. 

As an example of the above noted procedure, theoretical mini 
curves-of-growth  for seven selected H$_2$O blends on the spectrum of 
GJ\,229 are shown by the solid lines in Fig.\,4: The reference number 
of each blend (see Table\,3)
is shown for the corresponding mini CG, on which the filled circle
indicates the values of the observed log\,$(W/\lambda)_{\rm obs}$ (read on
the ordinate of Fig.\,4) and the derived abundance correction 
$\Delta{\rm log}\,A_{\rm O}$ (read on the abscissa). The resulting seven
values of $\Delta{\rm log}\,A_{\rm O}$ and their mean value are shown
by the filled circles and dashed line, respectively, in Fig.\,5d.

Some other examples of the resulting $\Delta{\rm log}\,A_{\rm O}$ 
values plotted against the observed values of log\,$(W/\lambda)_{\rm obs}$ are 
shown in Fig.\,5  for relatively early M dwarfs, in the order of the 
decreasing effective temperatures. The number of H$_2$O blends measured 
is rather small in the early M dwarfs such as GJ\,338B (Fig.5a) and 
GJ\,205 (Fig.\,5b), and this is because H$_2$O lines are still weak and 
difficult to measure (see Fig.\,2). For this reason, the scatter of
the data is rather large and accuracy of the results cannot be very high.
The number of H$_2$O blends measured increases for M dwarfs with lower 
$T_{\rm eff}$ values, and H$_2$O blends of modest strength in these 
M dwarfs can be measured fairly accurately. Especially, most of the 17 
selected blends in Fig.\,2 can be used in  M dwarfs such as GJ\,411 
(Fig.\,5i) and GJ\,436 (Fig.\,5j), and reasonably accurate results can 
be obtained for these cases.

More or less similar results are obtained in later M dwarfs shown 
in Fig.\,6. However, in the coolest end including  GJ\,777B (Fig.\,6h) 
and GJ\,406 (Fig.\,6j), 
the numbers of blends measured decrease. This should be because the blending 
is so severe that the line profiles of the H$_2$O blends cannot be 
defined well and the number of H$_2$O blends measured  has decreased.
Nevertheless, it appears that the mini curves-of-growth analysis
on the selected blends works reasonably well even for later M dwarfs in which
the continua are depressed by as much as 15 \% (subsection 4.3).
In Figs.\,5 \& 6, a remarkable feature is that the abundance corrections
$\Delta{\rm log}\,A_{\rm O}$  plotted against the observed values of 
log\,$(W/\lambda)_{\rm obs} $ show little systematic effect and 
distributed nearly horizontally. As a result,  the mean abundance 
correction can be well determined. 
We will discuss on this result further in subsection 6.1.

The logarithmic abundance correction $\Delta{\rm log}\,A_{\rm O}^{\rm A}$,
 the resulting logarithmic oxygen abundance
$$ {\rm log}\,A_{\rm O}^{\rm A} = {\rm log}\,A_{\rm O}^{0} + 
\Delta{\rm log}\,A_{\rm O}^{\rm A}, \eqno(3) $$
and the number of H$_2$O blends used for each M dwarf, $N_{A}$, are given
in the third, fourth, and fifth columns, respectively 
in Table\,6, following the object's name and 
the logarithmic carbon abundance $ {\rm log}\,A_{\rm C}$ from Paper I 
in the first and second columns, respectively. 

\vspace{2mm}
--------------------------------------------------------------------

Fig.\,4: Examples of mini curves-of-growth for H$_2$O blends
         in GJ\,229 (p.17). 

Fig.\,5: Derived values of the abundance correction 
         $\Delta\,{\rm log}\,A_{\rm O}^{A}$ from H$_2$O blends in the region A
         (early M) (p.18).

Fig.\,6: Derived values of the abundance correction 
         $\Delta\,{\rm log}\,A_{\rm O}^{A}$ from H$_2$O blends in the region A
         (late M) (p.18).

\vspace{2mm}

Table\,6: Mean values of the abundance correction 
         $\Delta\,{\rm log}\,A_{\rm O}^{A}$ and the resulting 
         ${\rm log}\,A_{\rm O}^{A}$ from H$_2$O blends in the region A (p.32). 

---------------------------------------------------------------------

\subsection{ Synthetic Spectra}

    We compare the observed spectra with the predicted ones based on the
oxygen abundances  $ {\rm log}\,A_{\rm O}^{\rm A}$ obtained from the
H$_2$O blends in the region A 
for all the M dwarfs in Table\,6, and some results are shown in
Fig.\,7. In the early M dwarf GJ\,338B, the observed H$_2$O blends
(filled circles) are rather weak but most of them can be reproduced well 
by the predicted spectra (solid line). For this reason, we are sure that we
have certainly identified H$_2$O features in this early M dwarf. 
In other early 
M dwarfs including GJ\,380, GJ\,820B, GJ\,884, and HIP\,12961, we can
recognize a few strong blends as due to H$_2$O, but other weak features
cannot be explained well as due to H$_2$O lines, and we decide to exclude
such cases from our analysis. In other 38 M dwarfs, the observed H$_2$O
blends are reproduced by the predicted spectra fairly well in general,
even if the matchings of the observed and predicted spectra are not perfect.
Since there may be some unknown blends other than H$_2$O and some noise due
to imperfect cancellations of the atmospheric lines, for example, the perfect
matching may anyhow be difficult. 
 
The candidates of H$_2$O blends used for analysis are reproduced from
Fig.\,2 and shown by the arrows in Fig.\,7c, but not all of these 
candidates are used in the actual analysis. For example, in the worst case
GJ\,777B, only five blends (log\,$(W/\lambda)_{\rm obs}$ given in Table\,3) 
are used in the analysis as in Fig.\,6h. In this case, the pseudo-continuum 
can be  defined (Fig.\,7f) but many H$_2$O features are so strong and
blendings are quite severe. For this reason,  the 
profiles cannot reach the pseudo-continuum level and thus are not well 
defined to meet our criterion to accept the blends for analysis. This 
result implies that it is difficult to analyze the spectrum composed of
the blends of many strong lines even if the pseudo-continuum can be
defined. For this reason, we will analyze H$_2$O blends in the region B
where the blendings are not so severe (section\,5).
         
We also evaluate $\chi^{2}$ values by
   $$ \chi^{2} = {1\over{N-1}}\sum_{i=1}^{N}
   {\Bigl(}{ {f_{\rm obs}^{i} - f_{\rm cal}^{i} }\over\sigma_{i}}
   {\Bigr)}^{2},   \eqno(4) $$
where $f^{i}_{\rm obs}$ and $f^{i}_{\rm cal}$ are the observed 
and predicted spectra, both  normalized by their pseudo-continua. 
$N$ is the number of
data points and $\sigma_{i}$ is the noise level obtained from the
$S/N$ ratio (Table\,1 of Paper I) assumed to be independent of $i$.
We  evaluate $\chi^{2}$ values for the predicted spectra based on
$ {\rm log}\,A_{\rm O}^{\rm A}$ and $ {\rm log}\,A_{\rm O}^{\rm A} 
\pm 0.05$ where  $ {\rm log}\,A_{\rm O}^{\rm A}$ 
is the oxygen abundances obtained above (fourth column of Table\,6) 
and 0.05 is about the probable error of the resulting 
$ {\rm log}\,A_{\rm O}^{\rm A} $.
The resulting values of $\chi^{2}$ for  $ {\rm log}\,A_{\rm O} =
 {\rm log}\,A_{\rm O}^{A}$ -0.05, +0.0, and +0.05 are shown
as $\chi^{2}_{-}$, $\chi^{2}_{0}$, and  $\chi^{2}_{+}$ in the 
sixth, seventh, and eighth columns, respectively, in Table\,6. 
We confirm that the $\chi^{2}_{0}$ value for $ {\rm log}\,A_{\rm O} =
 {\rm log}\,A_{\rm O}^{A}$ is smaller than those for
$ {\rm log}\,A_{\rm O} $ values deviating by about the probable error
from ${\rm log}\,A_{\rm O}^{A}$ for each object,
 except for a few cases. This result implies that our oxygen 
abundance ${\rm log}\,A_{\rm O}^{\rm A}$ is consistent with the spectral 
synthesis analysis in most cases.

We now examine the case that the $\chi^{2}$ value suggests a possible 
poor fitting. In the worst case GJ\,406, $\chi^{2}$ value is quite 
large and larger for our resulting oxygen abundance 
${\rm log}\,A_{\rm O}^{\rm A}$ than for reduced 
oxygen abundance (Table\,6). We show the observed and predicted spectra
for this M dwarf for a) log\,$A_{\rm O}$ = log\,$A_{\rm O}^{\rm A}$ 
($\chi^{2}$ = 13.204) and b) log\,$A_{\rm O}$ = log\,$A_{\rm O}^{\rm A}$ 
- 0.05 ($\chi^{2}$ = 10.510) in
Fig.\,8a and 8b, respectively. Also shown are the H$_2$O blends actually
used for the analysis  of this M dwarf by the arrows (see Table\,3). 
Inspection of Fig.\,8 reveals that the fittings of the H$_2$O blends
used for the analysis (those indicated by the arrows)
are in fact better in Fig.\,8a with larger $\chi^{2}$ value than in Fig.\,8b 
with smaller $\chi^{2}$ value. The reason why $\chi^{2}$ value is smaller
in Fig.\,8b should be due to better fittings in the features not used
in our analysis. To confirm this, we exclude the spectral region not used
for our analysis by masking the region between 20354.5 and 20385.2 {\AA} 
in our $\chi^{2}$  analysis (the region filled with yellow or grey in 
Fig.\,8). The result is $\chi^{2}$ = 11.119 and 11.227 for
${\rm log}\,A_{\rm O}$ = ${\rm log}\,A_{\rm O}^{\rm A}$ and
${\rm log}\,A_{\rm O}^{\rm A}$ -0.05, respectively. Thus, we are now convinced
that the result of the mini CG analysis is consistent with the 
$\chi^{2}$  analysis.

\vspace{2mm}
--------------------------------------------------------------------

Fig.\,7: Observed and predicted spectra of H$_2$O in the region A
         of six M dwarfs (p.19).

Fig.\,8: Observed and predicted spectra of H$_2$O in the region A
         of GJ\,406 (p.20).
         
--------------------------------------------------------------------
  
\subsection{The True- and Pseudo-Continua} 
The analyses so far are based on the spectra normalized by the 
pseudo-continua. This is because the true-continuum cannot be seen 
on the observed spectrum. In the predicted spectrum, the true-continuum 
is also not shown explicitly, but it can easily be evaluated by 
considering the continuous opacity alone in computation. The 
true-continuum levels obtained in this way and the predicted spectra 
normalized by the true-continua for GJ\,338A,
GJ\,15A, GJ\,611B, and GJ\,406 are shown in Fig.\,9 by the dotted and 
solid lines, respectively. The pseudo-continuous levels are also
shown by the dashed lines on the predicted spectra. The ratios of
pseudo- to true-continua, $F_{\rm pc}/F_{\rm tc}$, are estimated
to be 0.986, 0.950, 0.841, and 0.856 for the model spectra of GJ\,338A,
GJ\,15A, GJ\,611B, and GJ\,406, respectively.    

Although the true-continuum level cannot be seen in the observed spectrum, 
it can be estimated from the pseudo-continuum level in the observed 
spectrum by assuming the same $F_{\rm pc}/F_{\rm tc}$ ratio
for the observed spectrum as for the predicted spectrum. 
This is possible since the pseudo-continua for the observed spectra 
are defined well as in the predicted spectra in all the M dwarfs of
Fig.\,9. The observed 
spectra of GJ\,338A, GJ\,15A, GJ\,611B, and GJ\,406 normalized by 
the true-continua estimated in this way are shown by the filled circles 
connected by the  dotted lines in Fig.\,9. 
From this result, we now know that the pseudo-continua of the observed 
spectra in GJ\,338, GJ\,15A, G\,611B, and GJ\,406 are depressed by 1.4, 
5.0, 15.9, and 14.4 \%, respectively, from the true-continua.
     
\vspace{2mm}
--------------------------------------------------------------------

Fig.\,9: True- and pseudo-continua in four representative M dwarfs (p.21).

--------------------------------------------------------------------

\section{Analysis of  H$_2$O Lines in the Region B }

\subsection{Mini Curves-of-Growth Analysis of H$_2$O Blends}
In the region B, H$_2$O lines can be measured in 20 M dwarfs as mentioned 
in subsection 3.4, and we carry out the mini curve-of-growth analysis  
blend-by-blend as in subsection 4.1.  Some examples of the resulting 
$\Delta{\rm log}\,A_{\rm O}$ values plotted against the observed 
values of log\,$(W/\lambda)_{\rm obs} $ are shown in Fig.\,10 for the same 
dwarfs in Fig.\,6 for comparison. In general, considerable number 
of blends can be measured  in dM3.5 - dM4 dwarfs which are the 
major constituents of our sample. The number of blends measured 
is rather small in the dM2.5 dwarf GJ\,797B-NE and more difficult 
in M dwarfs with $T_{\rm eff} \gtrsim 3500$\,K.
Actually, we measure several possible H$_2$O blends in the M2 dwarf GJ\,250B 
($T_{\rm eff}$ = 3567\,K), but other predicted H$_2$O blends cannot be
definitively identified and we decide not to include such a case in our
analysis. On the other hand, even in M dwarfs with $T_{\rm eff} 
\lesssim 3500$\,K, it is not always possible to find sufficient number 
of H$_2$O lines to be analyzed. For example, we cannot obtain oxygen
abundances in GJ 411 ($T_{\rm eff}$ = 3465\,K) and GJ436 ($T_{\rm eff}$ 
= 3416\,K) from H$_2$O lines in the region B. We investigate the reason
for this and notice that the oxygen abundances in these stars are
about log\,$A_{\rm O} \approx -3.4$ (Table\,6), which are near the lower
limit in our M dwarfs (Fig.\,17). For this reason, H$_2$O lines are rather
weak in these objects. Thus, the detection of weak H$_2$O lines depends 
not only on temperature but also on oxygen abundance.      

  Even in the cooler dwarfs, the blendings are  not so severe
and the pseudo-continua are well defined. In fact, the depressions of the
continuum levels estimated by the way outlined in subsection 4.3 are  
3.7\% in the dM2.5 dwarf GJ\,797B-NE and  4.8\% even for our coolest 
sample GJ\,406. For this reason, measurements of the EWs are easier in 
the region B than in the region A, and a large number of lines can be 
measured for such cooler dwarfs including GJ\,611B, GJ\,777B, and 
GJ\,406 (Figs.\,10h-j).
Thus, the region B is definitely better than the region A for the analysis of
H$_2$O abundances in later M dwarfs. Unfortunately, however, the region B is
not useful for the early M dwarfs in which H$_2$O features are very weak.  
 
The value of logarithmic abundance correction 
$\Delta{\rm log}\,A_{\rm O}^{\rm B}$,
the resulting logarithmic oxygen abundance ${\rm log}\,A_{\rm O}^{\rm B}$, 
and number of H$_2$O blends in the region B used for the analysis, $N_{B}$,
are given in the fourth, fifth and sixth  columns, respectively, in Table\,7, 
following the object's name, the value of the logarithmic 
abundance correction $\Delta{\rm log}\,A_{\rm O}^{\rm A}$ and the resulting 
logarithmic oxygen abundance ${\rm log}\,A_{\rm O}^{\rm A}$
based on the analysis of the region A from Table\,6 in the first, 
second and third columns, respectively. 

\vspace{2mm}
--------------------------------------------------------------------

Fig.\,10: Derived values of the abundance correction 
         $\Delta\,{\rm log}\,A_{\rm O}^{B}$ from H$_2$O blends in the region B
         (for the same objects as in Fig.\,6) (p.22).

\vspace{2mm}

Table\,7: Mean values of the abundance correction 
         $\Delta\,{\rm log}\,A_{\rm O}^{B}$ and the resulting 
         ${\rm log}\,A_{\rm O}^{B}$ from H$_2$O blends in the region B (p.33). 

---------------------------------------------------------------------

\subsection{ Synthetic Spectra}
 Since the H$_2$O blends are sparsely distributed in the region B,
we examine a selected region (about one fourth of the region B) shown 
in Fig.\,11 where relatively large
number of H$_2$O blends are found together. The candidates of the
H$_2$O blends to be analyzed  are reproduced from Fig.\,3 and shown
by the arrows in Fig.\,11c, but again not all these candidates are 
actually analyzed as in the region A. In this region B, the central
depths of H$_2$O features are only about 5\,\% or so in the
earlier sample such as GJ\,797B-NE and GJ\,687, and never reach 
15\,\%  even in the latest M dwarf in our sample GJ\,406. The predicted 
spectra based on the values of $ {\rm log}\,A_{\rm O}^{\rm B}$  
shown by the solid lines generally reproduce the observed ones
shown by the filled circles but some weak features are rather
noisy. For such sparsely distributed H$_2$O blends, $\chi^{2}$ test
may not be useful.

\vspace{2mm}
--------------------------------------------------------------------

Fig.\,11: Observed and predicted spectra of H$_2$O in the region B
         of six M dwarfs (p.23).

--------------------------------------------------------------------

\subsection{Comparison of the Results from the Regions A and B }

The differences of the values of the logarithmic oxygen abundances 
${\rm log}\,A_{\rm O}$ based on the regions A and B
are given in the seventh column of Table\,7. The differences appear to be
less than the probable errors of the analysis in the region A as well as
in the region B in most cases.
  We also plot the values of the logarithmic 
oxygen abundances ${\rm log}\,A_{\rm O}$ based on the regions B
against those based on the region A in Fig.\,12 for 20 objects
for which analyses are done both in the regions A and B.
The results from the regions A and B show no systematic difference
and agree well in general. 

   The result outlined above implies that the analysis of spectra
whose continua are depressed by as much as 15\,\% can provide the
results not much different from the spectra whose continua are
depressed by less than 5\,\%. Thus, it is confirmed that the spectral analysis 
referring to the pseudo-continua works well, and spectra whose 
continua are depressed by molecular veil opacity can be analyzed
only if the pseudo-continua can be predicted accurately by the
use of proper molecular data and if the EWs can be measured well 
by referring to the pseudo-continua. 

\vspace{2mm}
--------------------------------------------------------------------

Fig.\,12: Comparison of the logarithmic oxygen abundances from the 
          regions A and B (p.24).

--------------------------------------------------------------------
 
\section{Discussion}

\subsection{H$_2$O as an Abundance Indicator of Oxygen}

The major difficulty in stellar abundance analysis is that the derived
abundance depends on model photosphere used. For example, the solar carbon 
and oxygen abundances are revised downward by about 50\,\% \citep{All02}
against the classical values (e.g., \cite{And89}). This revision is based 
on a new solar model photosphere referred to as the 3\,D model in which 
a time-dependent dynamical model for the convective zone has been 
incorporated. It is true that the classical 1\,D models are too simple for 
the real solar photosphere and further sophistication would certainly
be welcome. However, verification of the sophisticated models 
may be by no means easy. 

For comparison, we remember another major revision of solar abundance
in the past: The downward revision of iron abundance by about 50\,\%.
This revision has been achieved by the use of Fe II lines whose $gf$-values 
were measured accurately for the first time instead of the Fe I lines used
for a long time (\cite{Hol90}; \cite{Bie91}). In solar photosphere, most 
iron atoms are singly ionized  rather than neutral. For this reason, the  
Fe$^{+}$ abundance is relatively insensitive to the change of the 
physical condition in the line-forming region  and hence will not be
affected by imperfect modelling of the solar photosphere.
Thus, iron abundance could have been determined accurately
by the use of the Fe$^{+}$ lines for the first time, and the discrepancy
between the solar iron abundance and the meteoritic iron abundance
\citep{And89} could have been resolved. This result was further confirmed
by \citet{Gre99} who succeeded to reconcile abundance results from the FeI 
and FeII lines with the use of a new solar model photosphere. 
The abundance determination depends on many factors, of which 
model photosphere is a crucial one. For this reason, 
stellar abundance should better be determined from the species that 
consumes the larger portion of an element and we refer such species as 
the major species for the element.

In Fig.\,13, the abundances of some molecules are plotted
against the depth in four model photospheres of $T_{\rm eff}$ =
2800, 3200, 3600, and 4000\,K (the same models as used in Fig.\,1). 
Also, we plot
$$ y({\rm CO}) = P_{\rm CO}/(P_{\rm H} +　2 P_{\rm H_2} ),   \eqno(5)  $$
on logarithmic scale in Fig.\,13 by dotted lines. If almost all the carbon
atoms are in CO, $ {\rm log}\,y({\rm CO}) \approx {\rm log}\,A_{\rm C} 
\approx -3.4$. This expectation is satisfied in almost all the region of
our four models in Fig.\,13 and this fact confirms that CO is the major
species of carbon in all our cases. 
Thus the carbon abundance can be determined accurately from CO,
the major species of carbon,
as noted in Paper I. We also plot
$$ y({\rm H_2O}) = P_{\rm H_2O}/(P_{\rm H} +　2 P_{\rm H_2} ),   \eqno(6)  $$
by dashed lines. If almost all the oxygen atoms left after CO formation
are in H$_2$O, $ {\rm log}\,y({\rm H_2O}) \approx {\rm log}\,(A_{\rm O} 
 - A_{\rm C})$. Inspection of Fig.\,13 shows that this case is realized 
for cooler models of $T_{\rm eff}$ = 3200 and 2800\,K, and H$_2$O is the
major species of oxygen in these cases. We confirm that the same situation 
applies to the models of $T_{\rm eff} \lesssim 3400$\,K (as for details of the molecular abundances in these models, see the UCM database$^3$). 
In the model of $T_{\rm eff}$ = 3600, $ P_{\rm H_{2}O} \approx
P_{\rm OH}$  and H$_2$O is no longer the major species of oxygen 
in the models of $T_{\rm eff} \gtrsim 3600$\,K.
Thus, the oxygen abundances determined from H$_2$O may depend somewhat on
models in early M dwarfs of $T_{\rm eff} \gtrsim 3600$\,K. Since H$_2$O lines
are anyhow weak in the early M dwarfs, OH may also be used as another
abundance indicator of oxygen. Except for the early M dwarfs, 
H$_2$O is the major species of oxygen and can be an excellent 
abundance indicator of oxygen in M dwarfs.

Another advantage of H$_2$O as abundance indicator may be that its
thermal velocity is rather large because of its light weight. In
fact, thermal velocity of H$_2$O  at $T = $3000\,K is $\xi_{\rm th} =$
1.66\,km\,sec$^{-1}$. This is considerably larger than the micro-turbulent 
velocity we have assumed ($\xi_{\rm micro} =$1\,km\,sec$^{-1}$).
Since actual turbulent velocity in the photospheres of M dwarfs is 
not likely to be larger than the
value we assumed (see e.g., \cite{Bea06}), Doppler broadening of H$_2$O lines
should be dominated by the thermal velocity. This possibility may be
consistent with the result that the abundance corrections 
$\Delta {\rm log}\,A_{\rm O}$
plotted against log\,$(W/\lambda)_{\rm obs}$ in Figs. 5, 6, and 10
show little tilt or nearly horizontal as a mean, as noted in 
subsection 4.1.  If Doppler broadening
is dominated by the turbulence as in (super)giant stars, the abundance
corrections based on the saturated lines depend largely on the assumed
turbulent velocity while those based on the weak lines are nearly
independent of the assumed velocity, at least for the case of the unblended
single lines. In our case of the blended lines in which 
several hundreds of lines per 1\,{\AA} interval are overlapping, it is not 
known how EWs are influenced by the micro-turbulent velocity, and this
problem will be investigated further hopefully with higher resolution
spectra.

  Since the discovery that the turbulence plays dominant role in
the line broadening in stellar spectra \citep{Str34}, how to determine
the turbulent velocity has been one of the major problems in 
abundance analysis. In fact, one reason why the results of the
abundance analyses by different authors do not agree well in
general is due to differences of the turbulent velocities applied by
different authors. We might have  emphasized somewhat too much
on the effect of micro-turbulence in the analysis of CO in Paper
I even though CO is heavier than H$_2$O and hence the relative role
of the thermal velocity is not so large as in H$_2$O.
 Now, so far as the spectral analysis of H$_2$O in  M dwarfs is concerned,
the uncertainty due to the micro-turbulent velocity is rather minor
because of the larger thermal velocity,
and this fact can be regarded as an additional advantage of H$_2$O as an 
abundance indicator of oxygen in M dwarfs.           

\vspace{2mm}
--------------------------------------------------------------------

Fig.\,13: The major species of carbon and oxygen in the photospheres
          of M dwarfs (p.24). 

--------------------------------------------------------------------

\subsection{Oxygen Abundances in M Dwarfs}

  We take the weighted mean (with the numbers of lines used as weights)
of the oxygen abundances resulting from the
analyses of the regions A (section\,4) and B (section\,5) for 20 M dwarfs
for which the two regions have been analyzed. The results and
those from the region A alone (18 objects) are summarized in Table\,8 
together with the carbon abundances from Paper I. Also, the classical 
and more recent solar carbon and oxygen abundances by \citet{And89} and 
\citet{Asp09}, respectively, are included in Table\,8 for comparison.

The resulting values of log\,$A_{\rm O}$ are plotted against the values of 
log\,$A_{\rm C}$ in Fig.\,14. The larger filled circles are based on the
mean oxygen abundances from the regions A and B while the smaller filled 
circles from the region A alone. The two representative solar values are
shown by $\solar$ marks. The initial assumed values of log\,$A_{\rm O}$ = 
log\,$A_{\rm C}$ + 0.30 are represented by the solid line in Fig.\,14.
The oxygen abundances are systematically lower than the assumed initial
values in the carbon-rich M dwarfs. This tendency can be shown more clearly
by plotting the values of log\,$A_{\rm O}/A_{\rm C}$ (given in the fourth
column of Table\,8) against the values of log\,$A_{\rm C}$ as illustrated
in Fig.\,15. Clearly abundances of oxygen relative to carbon are
systematically smaller in M dwarfs with the larger carbon abundances.
This result agrees well with the result by \citet{Gus99} who have determined
carbon abundances in 80 F and G dwarfs (and oxygen abundances from
\cite{Edv93}) in the larger metallicity range with the use of
the forbidden [CI] line, which is expected to be less sensitive to
the photospheric structure and non-LTE effect compared to the other carbon
abundance indicators for F and G dwarfs.

The carbon abundance may approximately represent the metallicity usually 
measured by the iron abundance. Then, Fig.\,15 indicates that  the 
$A_{\rm O}/A_{\rm C}$ ratios are larger at lower metallicities and gradually
decrease in higher metallicities.  Thus, Fig.\,15 suggests 
different formation histories of carbon and oxygen in the
evolution of the Galaxy. After the large production of oxygen by 
Type II supernovae/hypernovae in the early Galaxy, 
the decrease of oxygen and carbon with increasing metallicity 
in the metallicity range of Fig.15  is due to Type I supernovae which 
produce more iron, but the decrease of carbon may be tempered by the 
contribution of carbon from AGB stars \citep{Nom13}. However, the actual site
of carbon production is still  unclear as discussed in detail by \citet{Gus99}.
In view of the advantage of CO and H$_2$O as abundance indicators of 
carbon and oxygen, respectively, as noted in the previous subsection, 
the systematic decrease of the O/C ratios at the higher metallicities 
found in M dwarfs would provide the best observational constraint on the 
formation of carbon and oxygen in the chemical evolution of the 
Galactic disk.

We have shown in Paper I (see its Fig.\,15) that the solar carbon abundance is 
atypical for its metallicity compared to the M dwarfs in the solar 
neighborhood if the recently downward revised solar value \citep{All02} is 
adopted. To examine the case of oxygen, we summarize the metallicities [Fe/H] 
= log\,$A_{\rm Fe}^{*}$ - log\,$A_{\rm Fe}^{\odot}$ from the literature
(\cite{Mou78}; \cite{One12}; \cite{Nev13}) 
 in Table 9. The iron abundances log\,$A_{\rm Fe}$ are obtained
assuming solar value of log\,$A_{\rm Fe}^{\odot}$ = -4.5 \citep{Asp09} 
and, together with our oxygen abundances, O/Fe ratios are obtained.
The resulting values of log\,$A_{\rm O}/A_{\rm Fe}$ are plotted against
[Fe/H] in Fig.\,16. Except for a few case, O/Fe ratios show a systematic 
increase towards lower [Fe/H], confirming the well known excess production 
of oxygen in metal-poor era. The two solar results
are  shown by the $\odot$ marks: The higher O/Fe case (marked with 1) 
is based on the classical solar abundance \citep{And89} and the O/Fe ratio 
follows the general trend shown by the M dwarfs. To the contrary, the 
lower O/Fe ratio (marked with 2) based on the recent downward revised 
oxygen abundance \citep{Asp09} does not follow 
the general trend shown by the M dwarfs, or the solar oxygen abundance is 
atypical for its metallicity compared with the nearby M dwarfs which is
typical unevolved stars in the Galactic disk.  
On the other hand, it was shown  that the solar abundances 
including oxygen are typical for its metallicity, compared to the elemental 
abundances in unevolved stars in the solar neighborhood \citep{Edv93}.
The classical solar oxygen abundance \citep{And89} is consistent with
this result while the downward revised result (\cite{All02}; \cite{Asp09})
 is contradicting with this result.

So far,  evolution of chemical elements in the Universe has been studied 
mostly from the abundance analyses of G type dwarfs including the halo
sample, and there should be some reasons for this choice. For example,
these stars are relatively well understood with the Sun as the proto-type 
compared to very hot and very cool stars and, especially,
many metallic lines used for abundance analysis can  be observed in 
their spectra. However, one drawback to use G type dwarfs as tracers of 
the chemical history of the Galaxy, is that the photospheres of these
stars may possibly be polluted by accretion of the metal-rich matter
through encounter with the interstellar gas clouds during the evolution 
of the Galaxy, as has been pointed out first by \citet{Yos81}. In this 
case, the accreted matter will not be diluted  well 
because of the shallow convective zone in the G dwarfs, and the photospheric 
abundances will suffer appreciable changes from the values at their birth. 
Recently, observational support for this possibility has been presented 
by showing that the metallicity in halo G dwarfs is indeed larger by 
about 0.2\,dex compared with that in K dwarfs of the same kinematic 
characteristics but having deeper convective zone \citep{Hat14}.
 Such a dilution effect may 
be most efficient in M dwarfs whose envelopes are wholly convective and 
the photospheric abundances may not be changed very much from  the values 
at their birth. For this reason, M dwarfs could be the best tracers of the 
Galactic chemical evolution if chemical analysis can be extended 
to halo M dwarfs in the near future, and M dwarfs will provide unique 
contribution to our understanding on the element formation in the early 
Universe. 

Finally, frequency distribution of oxygen abundances in 38 M dwarfs
is shown in Fig.\,17. The oxygen abundances log\,$A_{\rm O}$ in M dwarfs are
mostly between -3.5 and -3.0. On the other hand, a larger fraction
of M dwarfs hosting planet(s) (those noted by $\ddagger$ in Table\,8)
are oxygen-rich, which is consistent with the results for carbon in Paper I 
and for metallicity in F, G, K, and M stars (e.g., \cite{Fis05}; 
\cite{Joh09}).   

\vspace{2mm}
--------------------------------------------------------------------

Fig.\,14: The logarithmic oxygen abundances log\,$A_{\rm O}$ vs.
log\,$A_{\rm C}$ in 38 M dwarfs and the Sun (p.25). 

Fig.\,15: The values of log\,$A_{\rm O}/A_{\rm C}$ plotted against
          log\,$A_{\rm C}$ in 38 M dwarfs and the Sun (p.25). 

Fig.\,16: The values of log\,$A_{\rm O}/A_{\rm Fe}$ plotted against
          [Fe/H] in 38 M dwarfs and the Sun (p.25). 

Fig.\,17:  Frequency distribution of M dwarfs against log\,$A_{\rm O}$ (p.25).

\vspace{2mm}

Table\,8: Carbon and oxygen abundances in M dwarfs (p.34).

Table\,9: Oxygen-to-Iron  ratios in M dwarfs (p.34).

--------------------------------------------------------------------

\subsection{Method of Abundance Analysis}

 The spectra of cool stars including M dwarfs are apparently very
complicated, and abundance analysis has been deemed difficult in
general. We have done some trials in Paper I and in this paper 
to simplify the abundance analysis in M dwarfs, and we summarize
our attempts.
  
 In stellar abundance analysis, the so-called curve-of-growth (CG) method 
played an important role as detailed in the textbooks of classical 
astrophysics (e.g. \cite{Uns55}). In its refined form, the so-called universal
curve-of-growth, the growth of the equivalent width (EW) broadened by the
Doppler and damping effects  has been expressed in  unified curves
in dependence on the effective number of species producing the line.
With this curve-of-growth, the observed EW could be transformed to 
the effective number of species and hence to the abundance of element
producing the line. This method has widely been applied to the stellar
abundance determinations until the middle of the previous century.
Despite the simplicity and wide applicability of the CG method, however,
it has some drawbacks. For example, the effect of the
photospheric structure could not be taken into account because of
the simplified assumptions needed to compute EWs universally
(i.e. for all the types of stars). Also, applications are limited
to a single line with well defined profile  which can be described by
the Voigt profile.

 To relax the problems in the CG method noted above, the so-called spectral
synthesis (SS) method has been widely used in recent years. In this method,
any model photosphere can be used in computing the synthetic spectrum
including the blended features. Then the observed spectrum is
compared with the predicted one characterized by several parameters including
abundance,  which are to be determined from the best fit between the 
observed and predicted spectra.
Despite the wide use of the SS method, however,
the stellar abundance determinations by different authors are still not 
well converged to a definite result (e.g., \cite{Leb12}). One problem is
that the different line-broadenings such as the rotation, micro- and
macro-turbulence could not be separated on the synthetic spectrum.
In this respect, the CG method has a definite advantage
in that the EWs depend on the micro-turbulent velocity but not on the
other line broadening, and thus micro-turbulence could be
well separated from the other broadenings by the analysis of EWs with
 the CG method. Also, the SS as 
well as CG method is based on the prerequisite that the synthetic 
spectra as well as EWs are measured by referring to the true-continuum.   

  We believe that our analysis applied in subsection 4.1 has relaxed
all the restrictions involved in the analyses noted above. As to the
problem of continuum, we have noticed that the spectral analysis can 
be carried out by referring to the pseudo-continuum if it can be evaluated 
accurately on the theoretical spectrum, and recent progress in molecular 
database indeed makes such an approach possible (Paper I). 
Also, analysis is not limited to a single line but any blended features
can be used. What is important is that the observed and theoretical
spectral features are defined by exactly the same way by referring
to the same reference (i.e., either the true- or pseudo-continuum).
The abundance cannot directly be obtained from the observed EW of the 
spectral feature selected, but  can be obtained easily with the use of
a simple curve-of-growth. This CG is not necessary to cover the 
large range of EW as in the classical CG, but it should cover the restricted
portion of the CG around the observed EW. Such a mini CG can be prepared
theoretically by the use of EWs evaluated from the synthetic spectra
including  all the lines contributing to the spectral features to be 
analyzed. For this purpose, it is sufficient to generate three synthetic 
spectra for slightly different abundances, from which theoretical EWs are 
evaluated by exactly the same way as EW is  measured on the observed spectrum. 
Then abundance can be obtained for the observed EW from the three-point 
mini CG.  This analysis is repeated  line-by-line or blend-by-blend for 
all the spectral features measured. 

    The mini CG method can best be applied to H$_2$O for which the
pseudo-continuum level is also determined by H$_2$O itself. Then, the
mini CG analysis directly provides the oxygen abundance. For comparison, 
in the case of CO discussed in Paper I, the pseudo-continuum level is 
determined not by CO itself but by H$_2$O and also EWs of CO measured
include the contamination due to H$_2$O.  For this reason, derived 
carbon abundance includes the effect of  H$_2$O. Even though this 
effect is already included in the synthetic spectra from which EWs 
used for mini CG are evaluated, the result depends on the oxygen 
abundance assumed. For this reason, it is required to repeat the above 
processes by using the revised oxygen abundance if accurate result is 
required. In our Paper I, however,  this process was skipped since 
the blending H$_2$O lines are rather weak and will have little effect.

      The micro-turbulent velocity can be determined if necessary: For
this purpose, we repeat the mini CG analysis with  several assumed values of
the micro-turbulent velocity. The resulting abundance corrections from 
different lines (or blends)  depend little on the observed EWs only if 
the assumed  micro-turbulent velocity is correct. In our analysis of M 
dwarfs, this requirement is almost satisfied for almost all the cases 
(see Figs. 5, 6, \& 10) for the assumed micro-turbulent velocity, but 
this is because the thermal broadening dominates over the  turbulent 
broadening as noted in subsection 6.1. If the turbulent broadening is 
dominating, the micro-turbulent velocity is determined so
that the abundance corrections depend little on the observed EWs, and some
examples are shown for the case of red giant stars using unblended lines
(e.g., \cite{Tsu08}). It is still to be shown if the micro-turbulent velocity
can be determined from the blended spectral features by the method
outlined above. 
  
  We believe that our abundance analysis is essentially very simple
and flexible enough to be easily applied to complicated blended spectra 
with depressed continuum, and we hope to have shown that the abundance 
analysis of cool stars is not especially difficult compared to that 
of other types of stars..  

\subsection{H$_2$O in Stellar Spectra other than M Dwarfs}
  We have shown that H$_2$O molecule can be an excellent abundance indicator
of oxygen in M dwarfs. Then, we wonder if this is true for other types of stars
including  M type giants and supergiants which show H$_2$O lines in their
spectra.  For example, H$_2$O lines are prominent in the spectra of Mira type
variables for which detailed spectroscopic analysis has been done by
\citet{Hin79},
who showed that H$_2$O lines are composed of multiple components originating 
from different layers reflecting the complicated dynamical structures of their
atmospheres. Then H$_2$O lines are primarily useful as the probes of the
atmospheric structures of Mira type variables and abundance analysis would
be more difficult.

Spectra of water vapor were observed in red giants and supergiants other than 
Mira type variables, but  in the ways somewhat unexpected. For example, 
the announcement to have detected H$_2$O lines in the early M supergiant 
Betelgeuse ($\alpha$ Ori, M2Iab) by \citet{Woo64} was so unexpected 
at that time that it could  not be appreciated correctly for a long 
time \citep{Tsu00}. Also, H$_2$O lines were observed in the K giant 
Arcturus ($\alpha$ Boo, K2IIIp) with high resolution in the atmospheric 
window near 12\,$\micron$ \citep{Ryd02} 
and in the K giant Aldebaran ($\alpha$ Tau, K5III) with low resolution
by ISO \citep{Tsu01}. Such results are  difficult to understand
within the framework of the classical theory of stellar photospheres.
Further, in non-Mira M giant stars, H$_2$O spectra were observed  by ISO  
but showed excess absorption that could not be accounted for by the LTE model 
photospheres (e.g., \cite{Dec03}). Such problems must be solved before
H$_2$O molecule can be used as  an abundance indicator of oxygen in red 
giant and supergiant stars.

Probably the rarefied and extended atmospheres of high luminosity stars
are more complicated to be described by the classical theory of
stellar photospheres. Recent observations revealed that H$_2$O and CO
are in fact found  not in the photospheres but in the warm molecular layers 
extended out to $\approx 1.5\,R_{*}$  in M giant and supergiant stars
 (e.g., \cite{Ohn04}; \cite{Mon14}). Moreover, it was found recently
from the  aperture synthesis imaging of Betelgeuse that such molecular 
layers or clouds are  asymmetrically extended and show large scale 
time-dependent dynamical structure \citep{Ohn11}.
Thus, interpretation of H$_2$O  spectra in red giant and supergiant
stars would require  hybrid model atmosphere consisting  of components
with different dynamical characteristics, and H$_2$O  spectra will serve
primarily as probes of the outer atmospheric structure of
high luminous cool stars. 

\section{Concluding Remarks}

 We have shown in the present paper and in Paper I that CO and H$_2$O spectra 
of M dwarfs can be  excellent abundance indicators of carbon and oxygen, 
respectively, and carbon and oxygen abundances in dozens of M dwarf stars 
have been determined rather well.  This conclusion may appear to be 
somewhat unexpected  in view of the difficulties so far experienced in 
the attempts to analyze the spectra of M dwarfs. However, a new possibility 
of determining chemical abundances in cool dwarf stars has been opened by 
the recent progress as follows:
  
   First, the progress in observations is remarkable. For examples, infrared 
spectra of high quality can be obtained by the use of new infrared detectors, 
fundamental parameters such as the effective temperatures are determined 
from the directly measured angular diameters, and high precision photometric 
and astrometric data on M dwarfs  are available especially from the space 
observations. 

  Second, the traditional abundance analysis appears to be difficult in
M dwarfs whose continuum levels are depressed by the numerous molecular lines.
However, given that the pseudo-continuous levels defined by the numerous
molecular lines can be evaluated accurately with the use of the recent
molecular database, spectral analysis  referring to the pseudo-continuum 
results in an identical result with that referring to the true-continuum,
as we have shown in the present paper. 

  Third, abundance analysis depends on model photospheres, which are 
unfortunately by no means perfect yet because of the extreme complexity 
of the real photospheres. For this reason, it is useful to do abundance 
analysis to be free from imperfect modelling of the photospheric structure 
as far as possible. In the photospheres of  M dwarfs, 
CO abundance is identical with the abundance of carbon itself in almost all
the M dwarfs and H$_2$O abundance is almost the same with 
$A_{\rm O} - A_{\rm C}$ except for early M dwarfs (subsection 6.1). 
Then, CO and H$_2$O abundances depend little on the model applied. 
For this reason, carbon and oxygen abundances can best be determined in 
M dwarfs rather than in other types of stars.   
 
   Thanks to the favorable conditions noted above, the abundance 
analysis of M dwarfs based on CO and H$_2$O has been done rather well 
within the framework of the classical LTE  analysis.
This shows a marked contrast to the case of molecular spectra in high 
luminous cool stars including K - M giant and supergiant stars, in which
classical LTE analysis encountered serious difficulties (subsection 6.4). 
As expected, LTE analysis is applicable to the high density and compact
photospheres of low luminous stars rather than to the low density and
extended atmospheres of high luminous stars. 
Although we certainly hope that the analysis of molecular spectra can be 
done consistently both in low and high luminous stars in the near future,
we now take the advantage that the spectra of M dwarfs can be analyzed on 
the basis of the well established classical theory of spectral line 
formation, and we hope that M dwarfs will be used more widely as the 
probes of the chemical abundances in the Universe.

\bigskip

We thank an anonymous referee for careful reading of the text, and
for providing many helpful suggestions that contributed to clarification
and better presentation of our result.

This research makes use of data products from the Wide-field Infrared
Survey Explorer which is a joint project of the University of
California, Los Angeles, and the Jet Propulsion Laboratory /
California Institute of Technology, funded by NASA.

This research has made use of the VizieR catalog access tool and
the SIMBAD database, both operated at CDS, Strasbourg, France, and
of the RECONS database in www.recons.org.

Computations are carried out on common use data analysis
computer system at the Astronomy Data Center, ADC, of the National
Astronomical Observatory of Japan.

\appendix

\section{ The AllWISE vs. the WISE All-Sky Catalogs}
The absolute magnitudes at 3.4\,$\micron$, $M_{\rm 3.4}$, based on the 
WISE $W1$ fluxes given in the AllWISE Catalog are shown in Tables 10 and 11,
while those in the  WISE All-Sky Release were shown in Tables 3 and 4 
of Paper I. The $M_{3.4} - {\rm log}T_{\rm eff}$ plots based on the data 
in Tables 10 and 11 are illustrated  in Fig.\,18 and the dashed line 
is the exact copy from  that given in Fig.\,1 of Paper I. As noted in 
section 2, the dashed line fits the plots in Fig.\,18 quite well, 
and can be applied to the data of both the AllWISE and 
the WISE All Sky Catalogs.

\vspace{2mm}
--------------------------------------------------------------------

Fig.\,18: The absolute magnitude $M_{\rm 3.4}$ based on the AllWISE
Catalog plotted against log\,$T_{\rm eff}$ (p.26).

\vspace{2mm}

Table\,10: $M_{\rm 3.4}$ and $T_{\rm eff}$ based on the interferometry
            (p.35).

Table\,11: $M_{\rm 3.4}$ and $T_{\rm eff}$ based on the infrared
flux method (p.35). 

--------------------------------------------------------------------

\newpage

\onecolumn

\begin{figure}
   \begin{center}
       \FigureFile(100mm,150mm){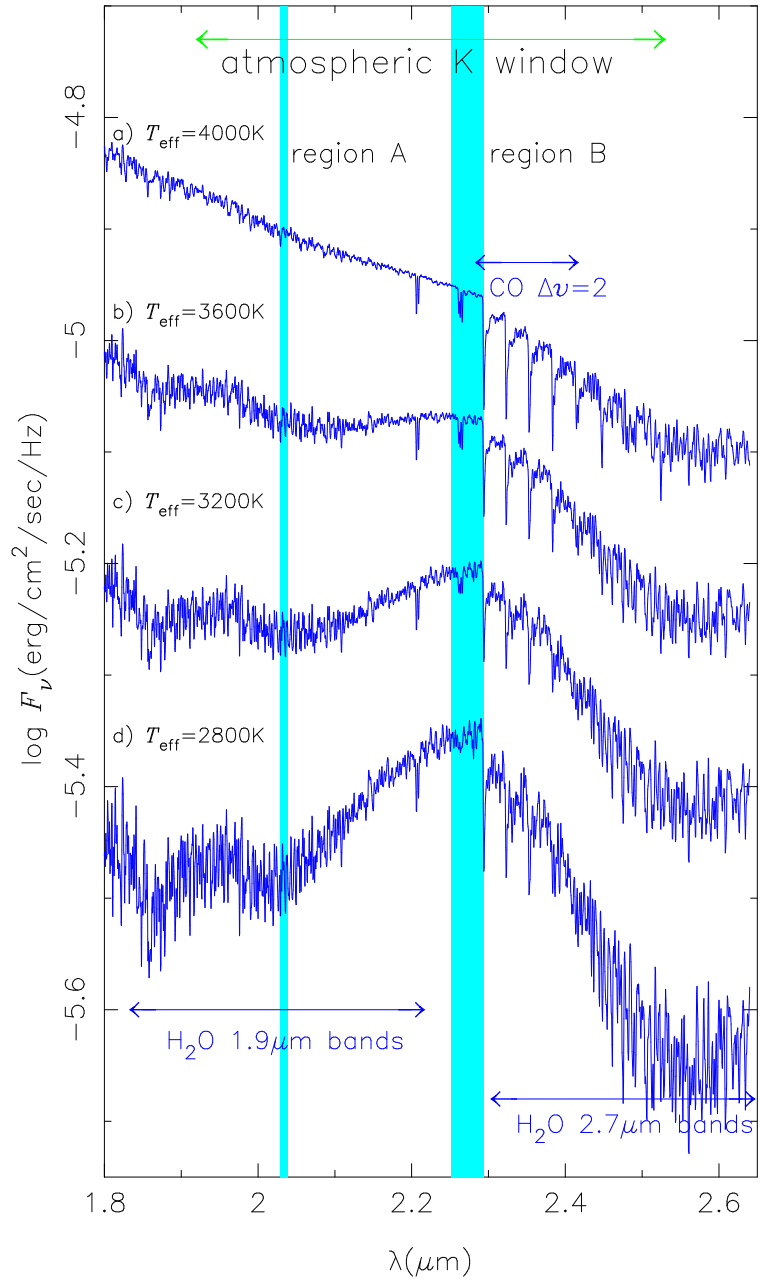}
   \end{center}
   \caption{Spectra of M dwarfs ($R = 2000$) in the $K$ band region
based on the model photospheres of 
$T_{\rm eff}$/log\,$g$ = a) 4000\,K/4.5, b) 3600\,K/4.75, c) 3200\,K/5.0, 
and d) 2800\,K/5.25. The H$_2$O 1.9 and 2.7\,$\micron$ bands are
increasingly stronger in cooler models, but the CO second overtone bands are
rather masked by H$_2$O bands in the cooler models. The regions A and B,
which are in the atmospheric window shown at the top of the figure,
are selected for detailed analysis of H$_2$O lines. The region A is
in the midst of H$_2$O 1.9\,$\micron$ bands while the region B 
suffers the effects of both the H$_2$O 1.9 and 2.7\,$\micron$ bands
least.      
}\label{figure1}
\end{figure}

\begin{figure}
   \begin{center}
       \FigureFile(160mm,160mm){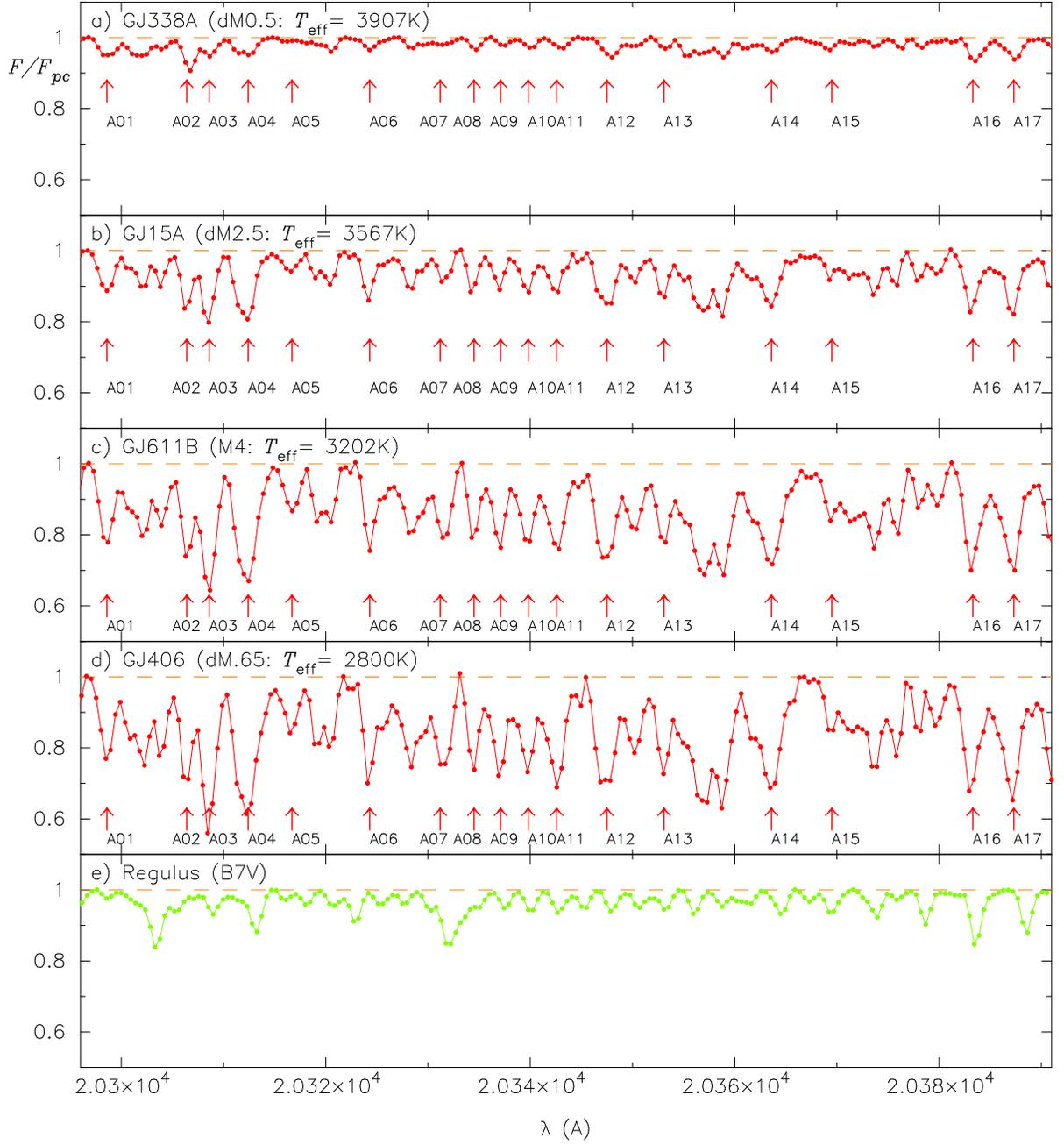}
   \end{center}
   \caption{The spectra of M dwarfs in the region A roughly in the order
 of the decreasing effective temperatures: a) GJ\,338A, b) GJ\,15A,
c) GJ\,611B, and d) GJ\,406. Also shown is e) Regulus used for calibration.
 The candidates of H$_2$O blends for detailed analysis are indicated
by arrows with the reference numbers A01--A17 to Tables\,2 \& 3.
}\label{figure2}
\end{figure}

\begin{figure}
   \begin{center}
       \FigureFile(160mm,160mm){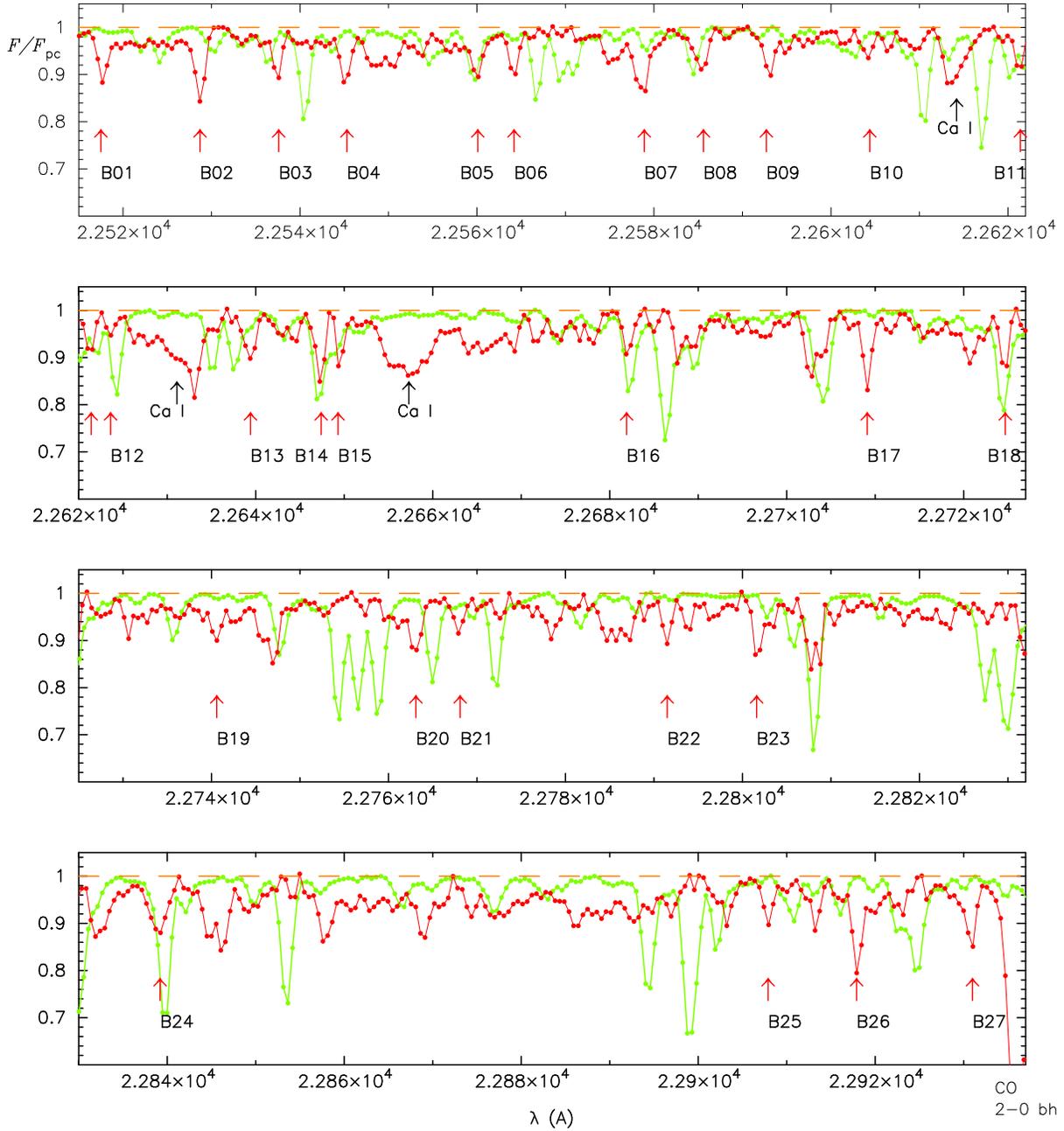}
   \end{center}
   \caption{The  spectrum of GJ\,406 (red or black) as an example of the
spectrum in the region B. The spectrum of Regulus (green or grey) is used
as the calibrator of the wavelength and the atmospheric absorption. 
The candidates of H$_2$O blends for detailed analysis are indicated
by arrows with the reference numbers B01--B27 to Tables\,4 \& 5.
The strong lines in Regulus are CH$_4$ and N$_2$O of telluric origins. 
}\label{figure3}
\end{figure}

\begin{figure}
   \begin{center}
       \FigureFile(100mm,120mm){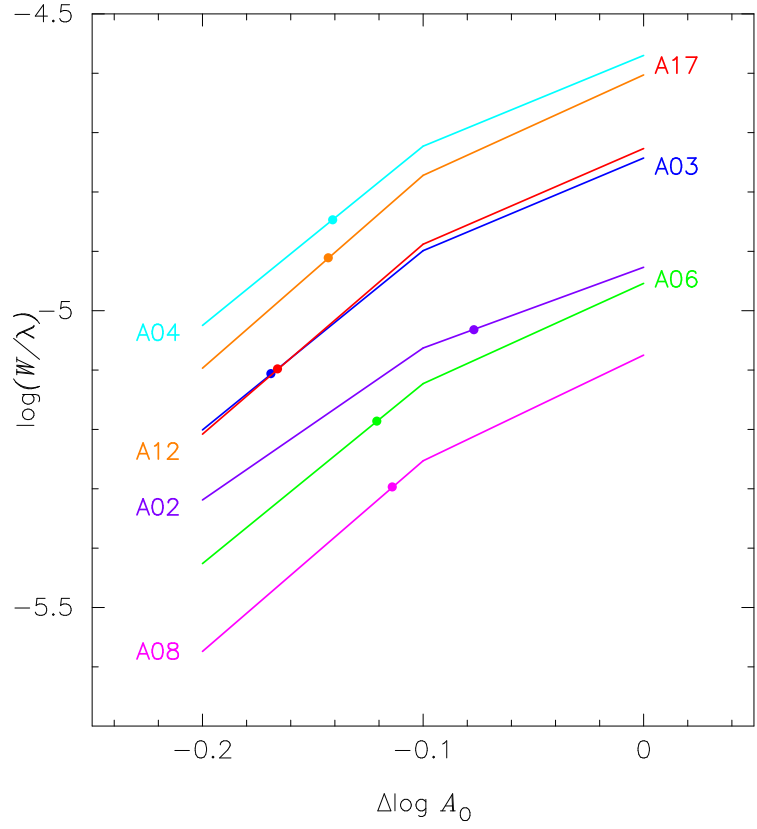}  
   \end{center}
   \caption{ Mini curves-of-growth for seven selected H$_2$O blends in 
GJ\,229, and the reference numbers to Table\,3 are shown for the 
corresponding mini CG.
The filled circle on each mini curve-of-growth indicates the observed
value of log\,$(W/\lambda)_{\rm obs}$ (from Table\,3 and read on the 
ordinate) and the 
resulting value of the abundance correction $\Delta{\rm log}\,A_{\rm O}$
(read on the abscissa and plotted on Fig\,5d by filled circle).   }
\label{figure4}
\end{figure}

\twocolumn

\begin{figure}
   \begin{center}
       \FigureFile(75mm,200mm){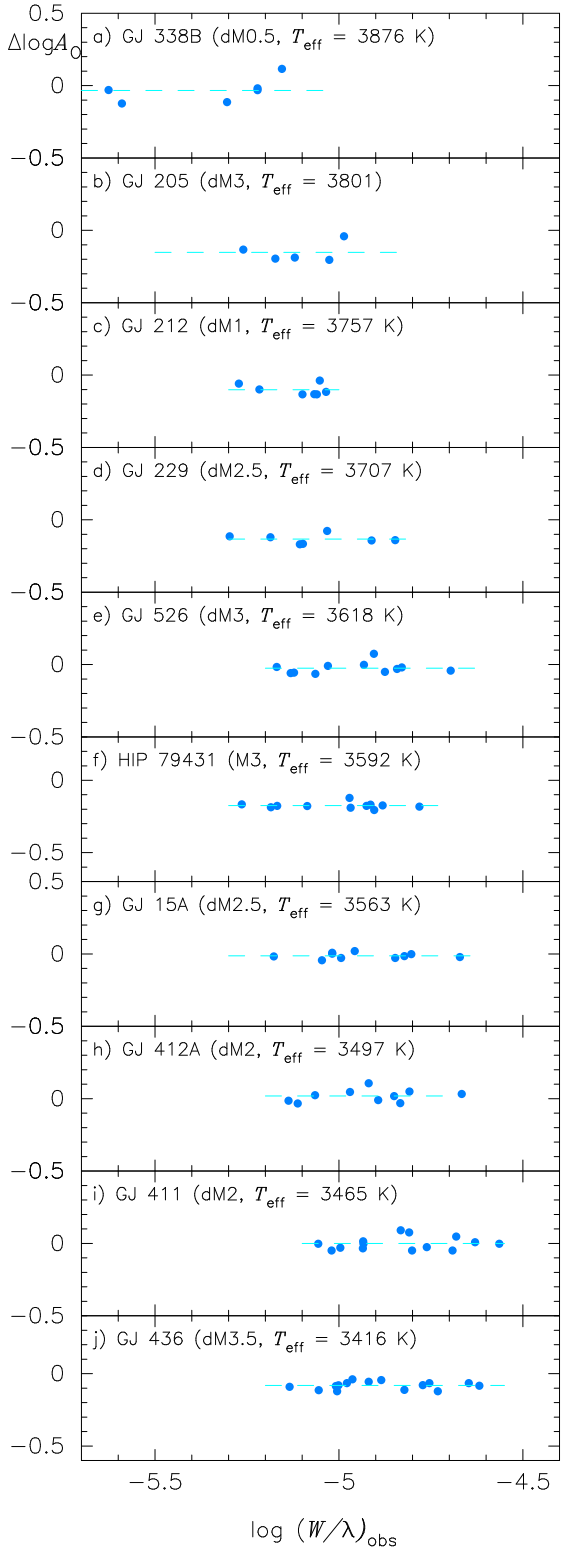}
   \end{center}
   \caption{The resulting logarithmic abundance corrections 
$\Delta{\rm log}\,A_{\rm O}$ by the mini curves-of-growth analysis for the 
H$_2$O blends in the region A are plotted against the observed values of
log\,$(W/\lambda)_{\rm obs}$. The dashed line shows the mean value of
$\Delta{\rm log}\,A_{\rm O}$. In this figure, relatively early M dwarfs
are shown in the order of decreasing effective temperatures.    }
\label{figure5}
\end{figure}

\begin{figure}
   \begin{center}
       \FigureFile(75mm,200mm){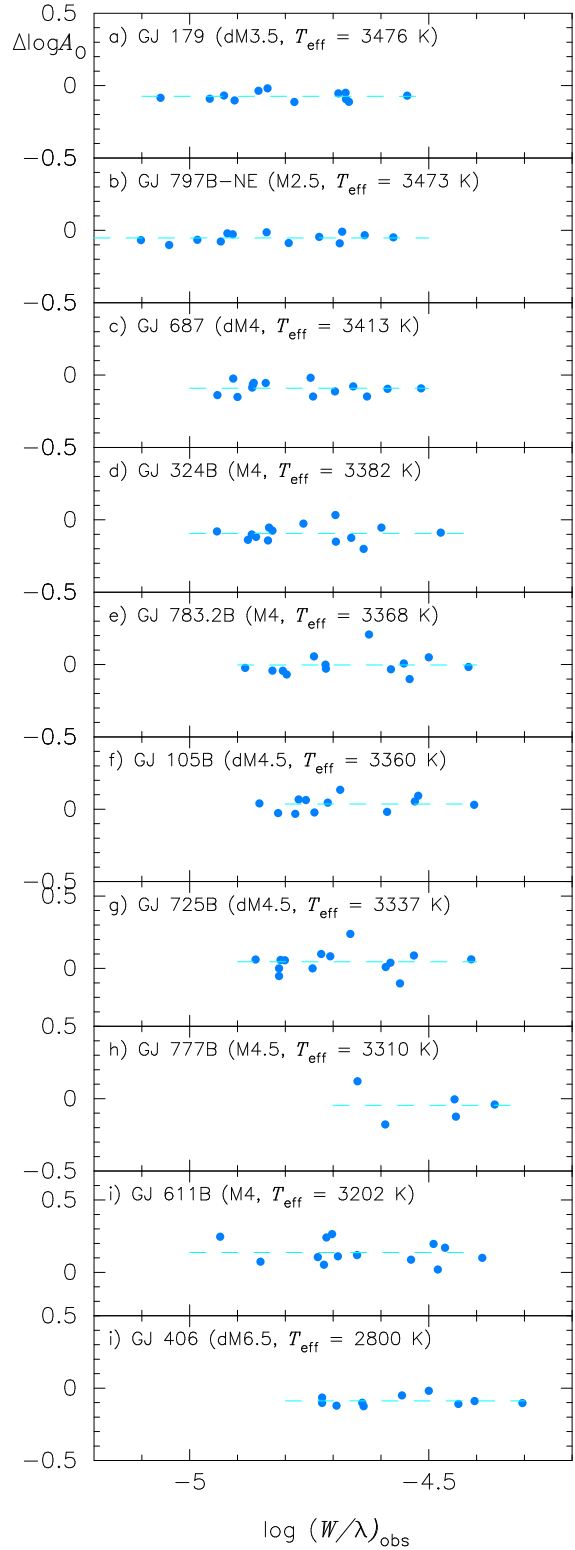}
   \end{center}
   \caption{The same as in Fig.\,5 but for later M dwarfs
in the order of decreasing effective temperatures.
    }
\label{figure6}
\end{figure}

\onecolumn

\begin{figure}
   \begin{center}
       \FigureFile(160mm,200mm){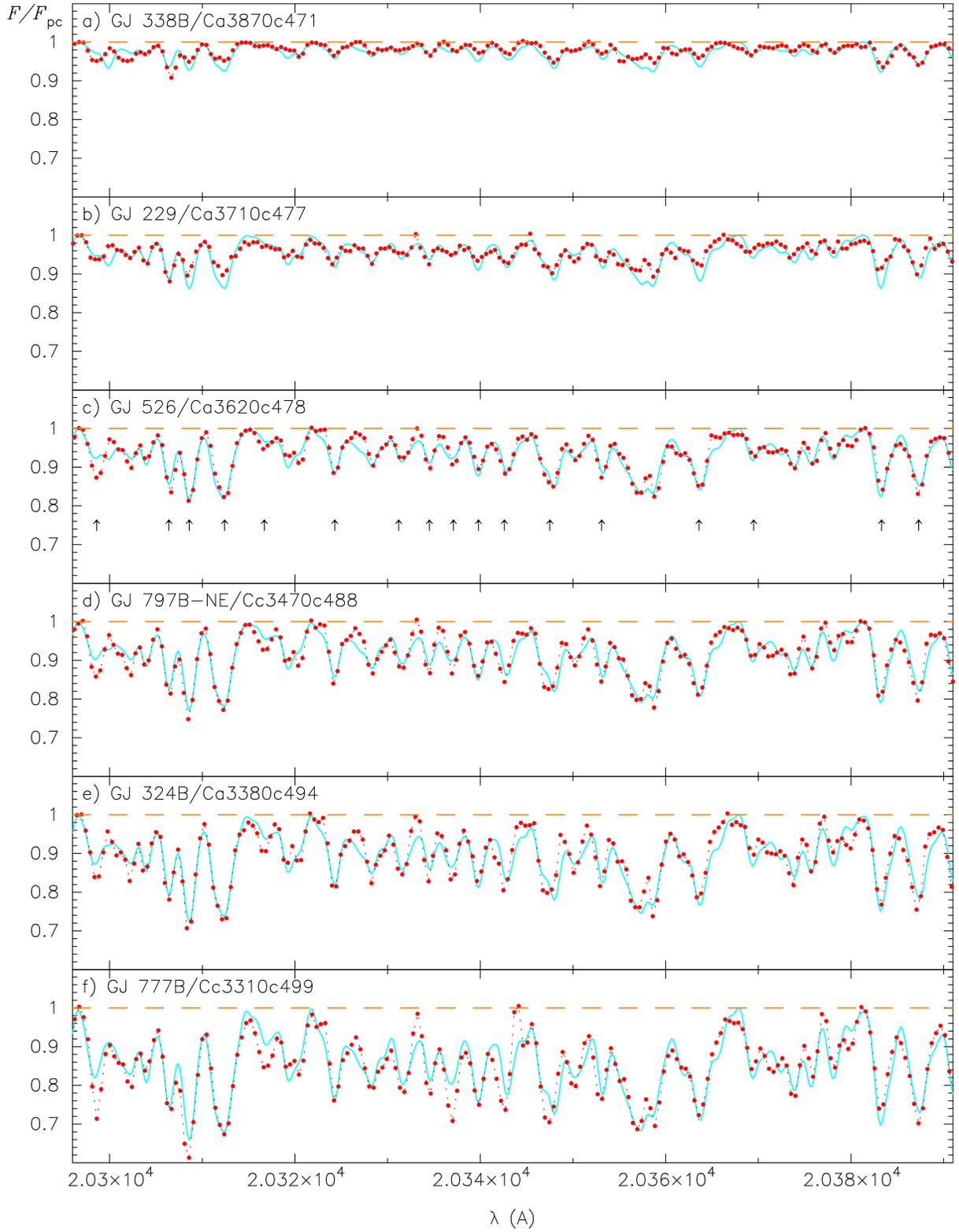}
   \end{center}
   \caption{
    Comparison of the observed (filled circles) and predicted (solid lines)
spectra based on the oxygen abundances log\,$A_{\rm O}^{\rm A}$ in Table\,6  
for six selected M dwarfs:
a) GJ\,338B (dM0.5) .
b) GJ\,229 (dM2.5).
c) GJ\,526 (dM3).
d) GJ\,797B-NE (M2.5).
e) GJ\,324B (M4).
f) GJ\,777B (M4.5).
 }
\label{figure7}
\end{figure}

\begin{figure}
   \begin{center}
       \FigureFile(160mm,75mm){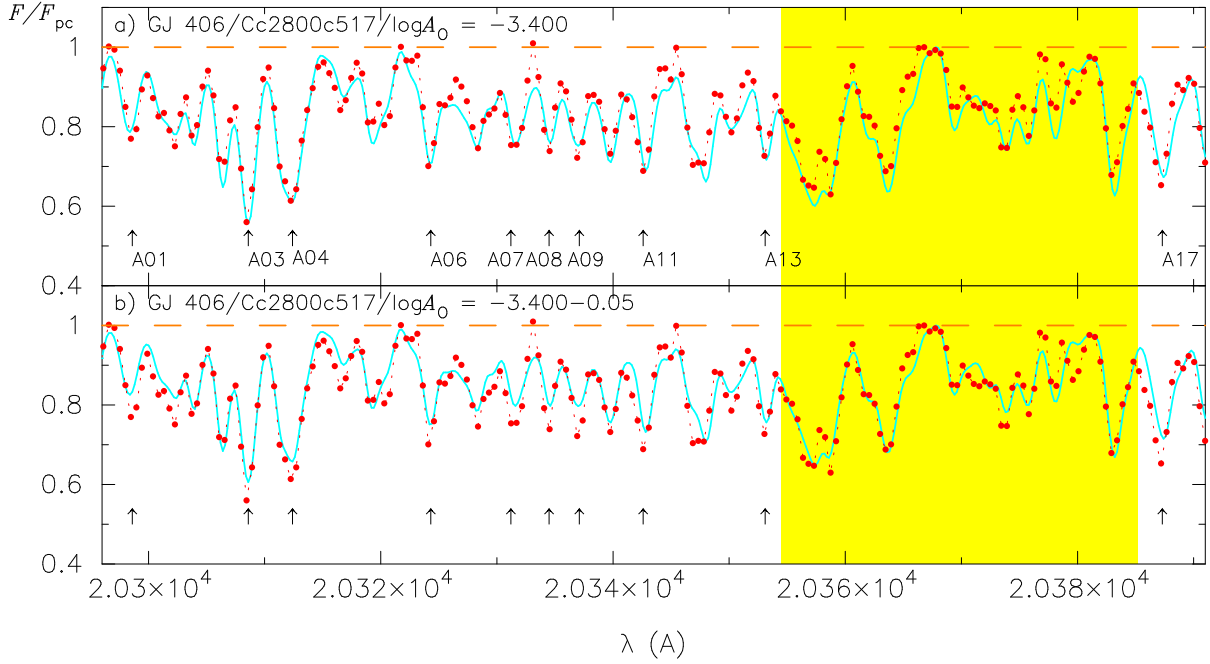}
   \end{center}
   \caption{Comparison of the observed (filled circles) and predicted 
(solid lines) spectra   for GJ\,406:
a) For the oxygen abundance log\,$A_{\rm O}$ = -3.40 in Table\,6 resulting
 $\chi^{2}$ = 13.204, but fittings are fine for the blends used for the
analysis (indicated by the arrows).
b) For the oxygen abundance log\,$A_{\rm O}$ = -3.40 -0.05, resulting
 $\chi^{2}$ = 10.510, but fittings are worse for the blends used for the
analysis (indicated by the arrows). Then the part of the spectrum not 
used for the analysis (between 20354.2 and 20385.2{\AA}; filled by yellow
or grey in the figure) is excluded from the $\chi^{2}$ test. The results are
 $\chi^{2}$ = 11.119 and $\chi^{2}$ = 11.227 for a) log\,$A_{\rm O}$ = -3.40
and b) log\,$A_{\rm O}$ = -3.40 -0.05, respectively, confirming that
the mini CG analysis and the $\chi^{2}$ analysis are consistent.
  }\label{figure8}
\end{figure}

\begin{figure}
   \begin{center}
       \FigureFile(160mm,95mm){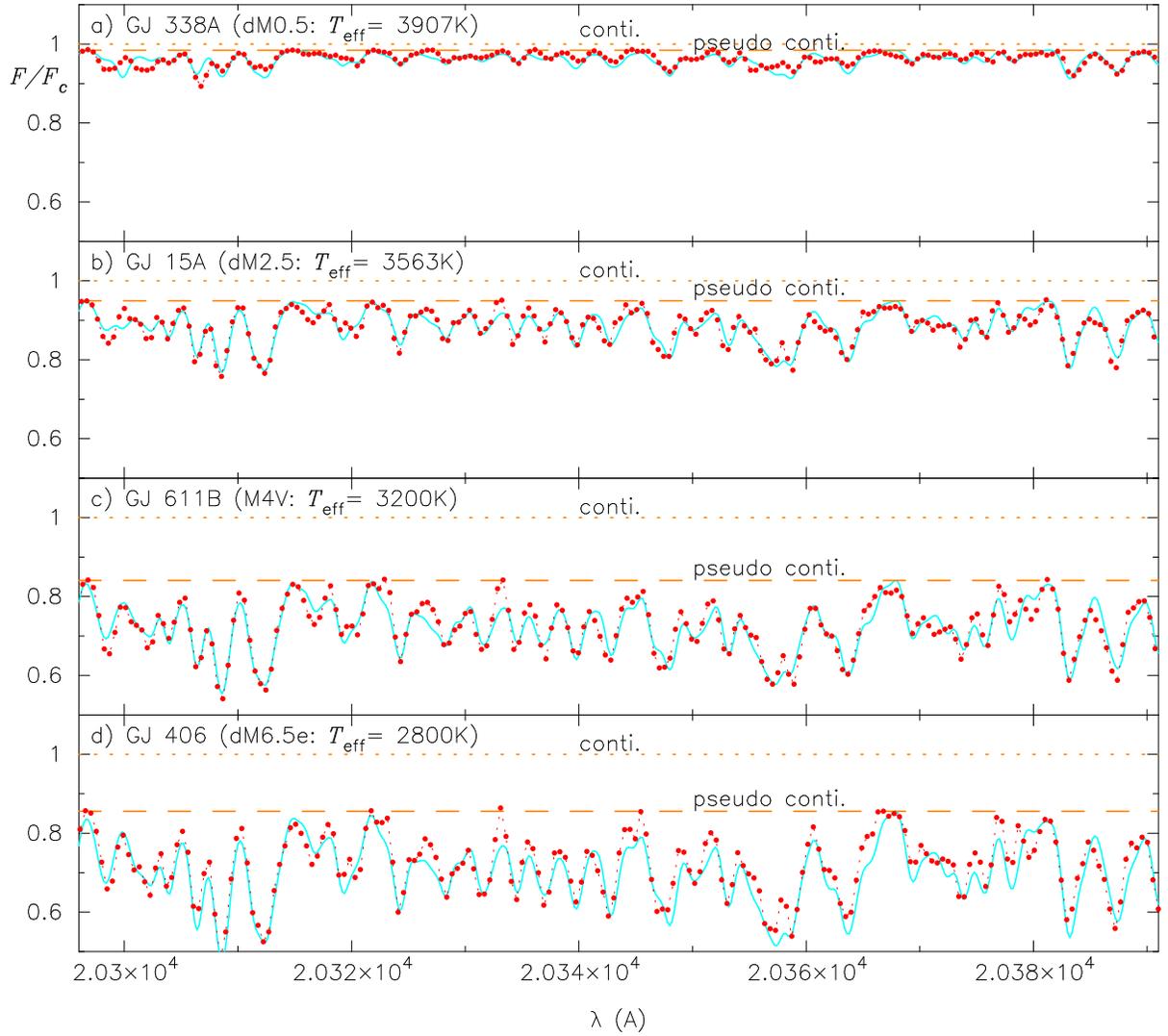}
   \end{center}
   \caption{ 
 Model spectra normalized by their true-continua shown
by the dotted lines are shown by the solid lines for four model spectra.
The observed spectra of the four stars normalized by their pseudo-continua
(dashed lines) in Fig.\,2 are shown here re-normalized by their 
true-continua estimated by applying the pseudo-to-true continua ratios of 
the model spectra. In this way, the depressions of the continuum levels 
of the four stars are estimated even though the true-continua can never be 
seen on the observed spectra.  
 }
\label{figure9}
\end{figure}

\begin{figure}
   \begin{center}
       \FigureFile(75mm,200mm){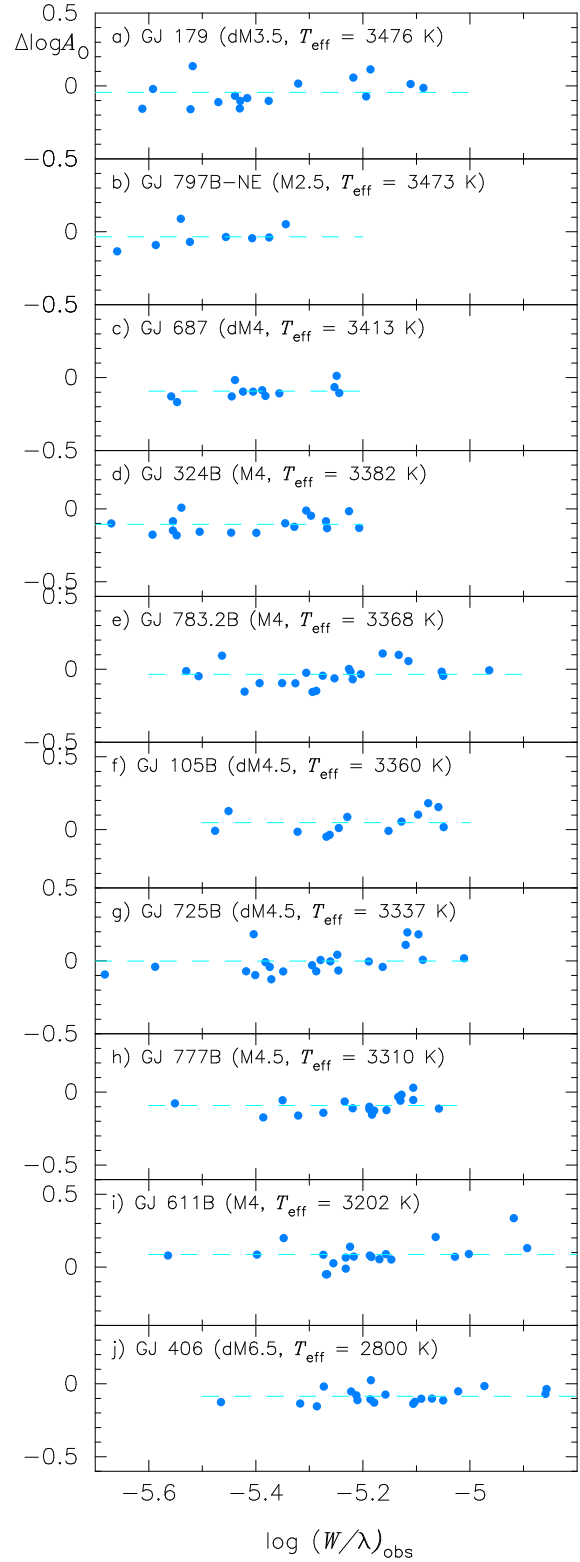}
   \end{center}
   \caption{The resulting logarithmic abundance corrections 
$\Delta{\rm log}\,A_{\rm O}$ by the mini curves-of-growth analysis 
for the H$_2$O blends in the region B plotted against the observed values of
log\,$(W/\lambda)_{\rm obs}$. The dashed line shows the mean value of
$\Delta{\rm log}\,A_{\rm O}$.
Compared with Fig.\,6 for the same objects, larger numbers of blends can be
used especially for later types.    }
\label{figure10}
\end{figure}

\begin{figure}
   \begin{center}
       \FigureFile(160mm,200mm){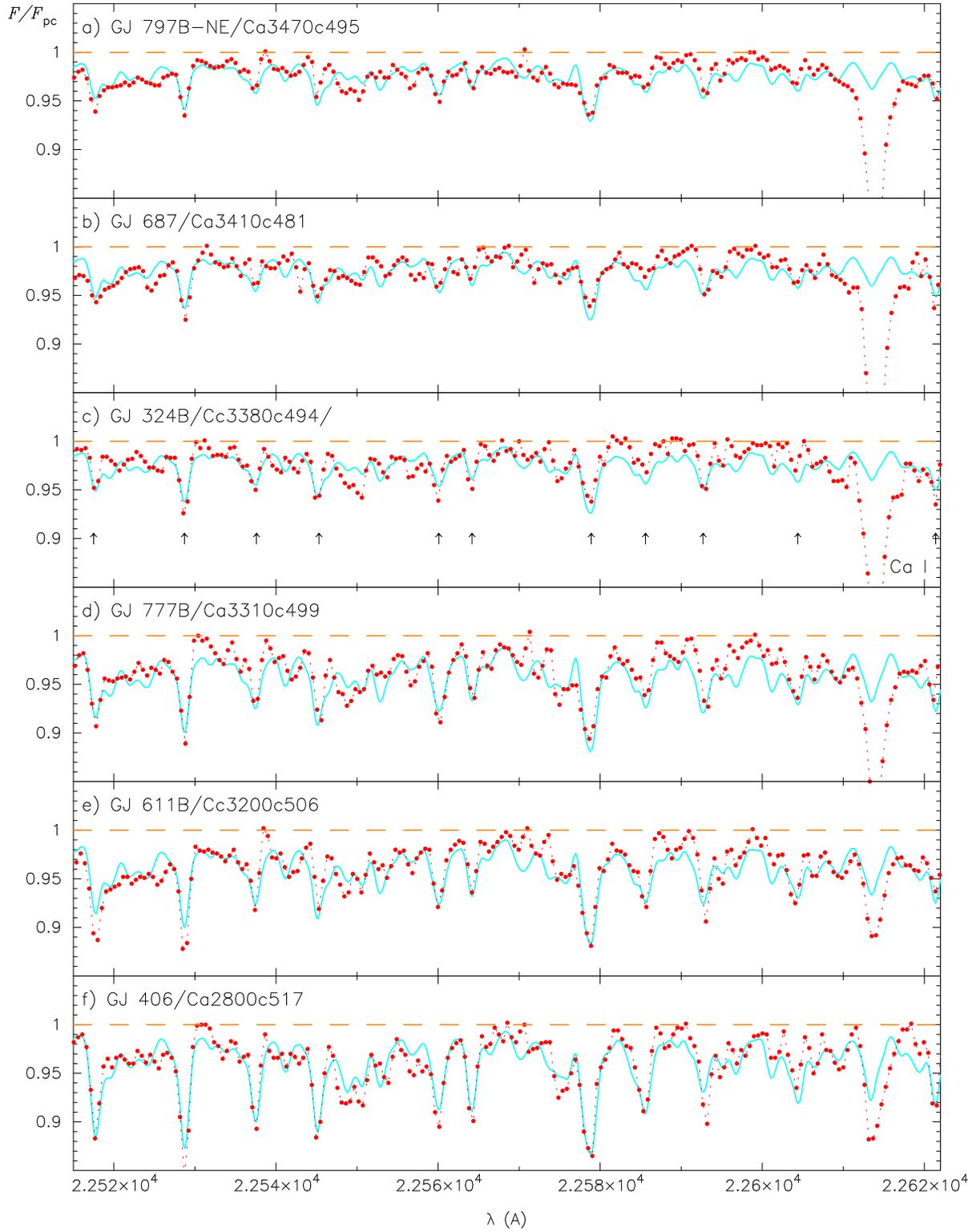}
   \end{center}
   \caption{ 
    Comparison of observed (filled circles) and predicted (solid lines)
spectra based on the oxygen abundances determined from H$_2$O blends in the 
region B  for six M dwarfs:
a) GJ\,797B-NE (M2.5). 
b) GJ\,687 (dM4).
c) GJ\,324B (M4). 
d) GJ\,777B (M4.5).
e) GJ\,611B (M4).
f) GJ\,406 (dM6.5).
The strong feature at 22614\,{\AA} in each observed spectrum is due to Ca\,I
line. But this Ca\,I line is not included in our line-list and the
feature at the position of the Ca\,I line on the predicted spectra
is not due to Ca but to H$_2$O.
}
\label{figure11}
\end{figure}

\twocolumn

\begin{figure}
   \begin{center}
       \FigureFile(80mm,80mm){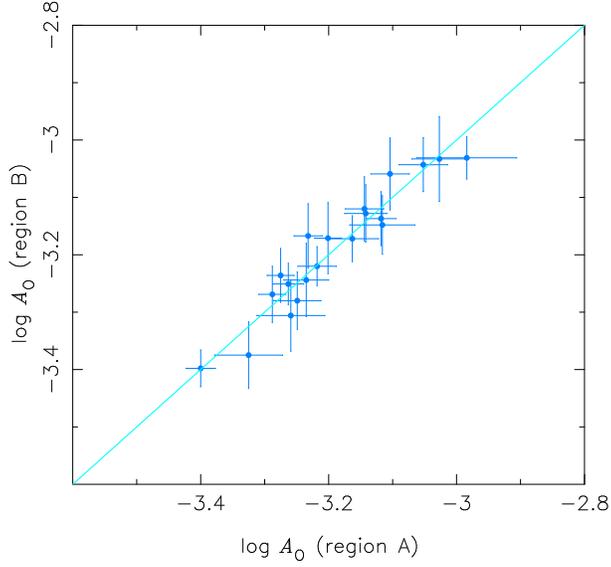}
   \end{center}
   \caption{ Oxygen abundances determined from the region B are plotted 
against those from the region A. The results from the different regions
agree  within the probable errors in most objects.
 }
\label{figure12}
\end{figure}

\begin{figure}
   \begin{center}
       \FigureFile(80mm,120mm){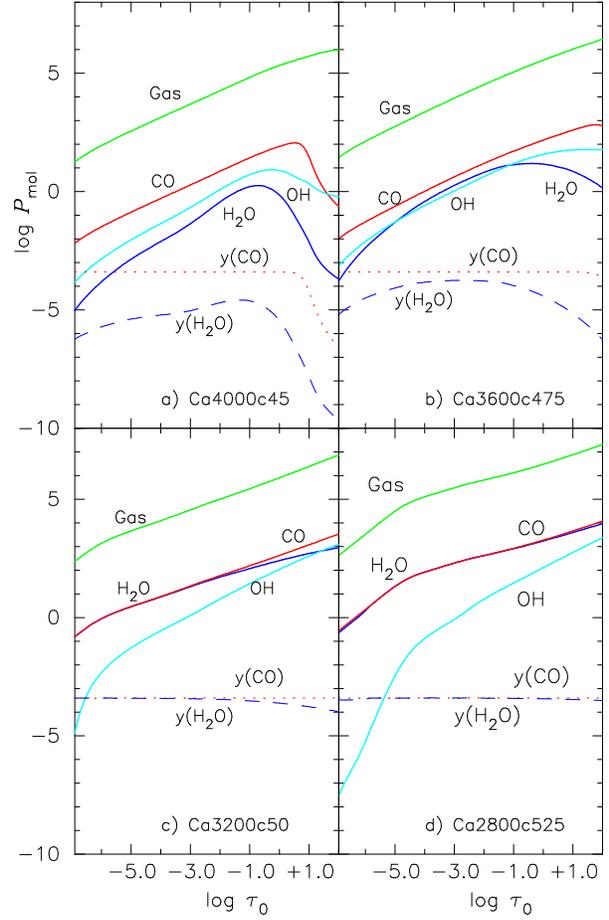}
   \end{center}
  \caption{
  Molecular abundances plotted against the optical depths in four model
photospheres of $T_{\rm eff}$/log\,$g$ = a) 4000\,K/4.5, b) 3600\,K/4.75, 
c) 3200\,K/5.0, and d) 2800\,K/5.25. Also, $ y({\rm CO})$ and 
$ y({\rm H_2O})$ are defined by equations (5) and (6), respectively, 
in the text. If almost all the carbon atoms are in CO (i.e., if CO is
the major species of carbon), $ y({\rm CO}) \approx A_{\rm C}$. If almost 
all oxygen atoms left after CO formation are in H$_2$O (i.e., if H$_2$O
is the major species of oxygen), $ y({\rm H_2O}) \approx A_{\rm O} - 
A_{\rm C}$. It can be confirmed that CO is the
major species of carbon in all the models and H$_2$O in the models of
$T_{\rm eff} \lesssim 3400$\,K. 
}
\label{figure13}
\end{figure}

\begin{figure}
   \begin{center}
       \FigureFile(80mm,80mm){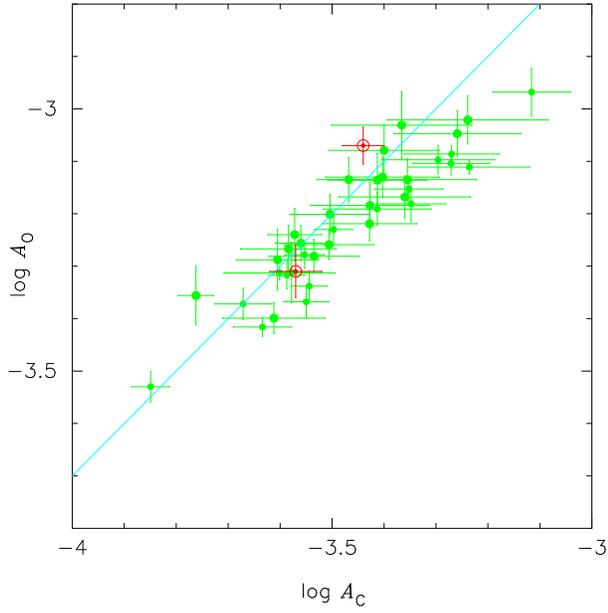}
   \end{center}
   \caption{The oxygen abundances log\,$A_{\rm O}$ are plotted against the 
carbon abundances log\,$A_{\rm C}$ in 38 M dwarfs. The classical high 
solar case is indicated by the upper $\odot$ and the recent 
downward revised solar case by the lower $\odot$. The assumed initial
relation of log\,$A_{\rm O}$ = log\,$A_{\rm C}$ + 0.30 is shown by the
sold line.
    }
\label{figure14}
\end{figure}

\begin{figure}
   \begin{center}
       \FigureFile(80mm,80mm){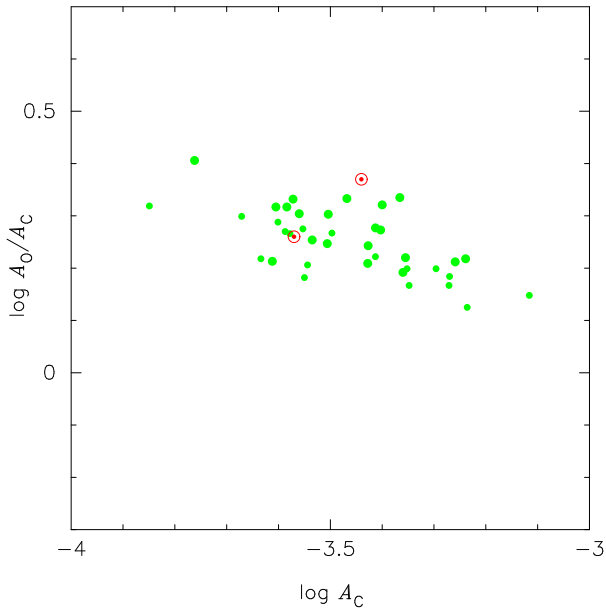}
   \end{center}
   \caption{The oxygen-to-carbon ratios are plotted against the carbon 
abundances log\,$A_{\rm C}$ in 38 M dwarfs, and in the Sun.
The oxygen-to-carbon ratios are smaller at the higher carbon abundances,
showing that carbon productions are more effective at higher metallicities.  }
\label{figure15}
\end{figure}

\begin{figure}
   \begin{center}
       \FigureFile(80mm,80mm){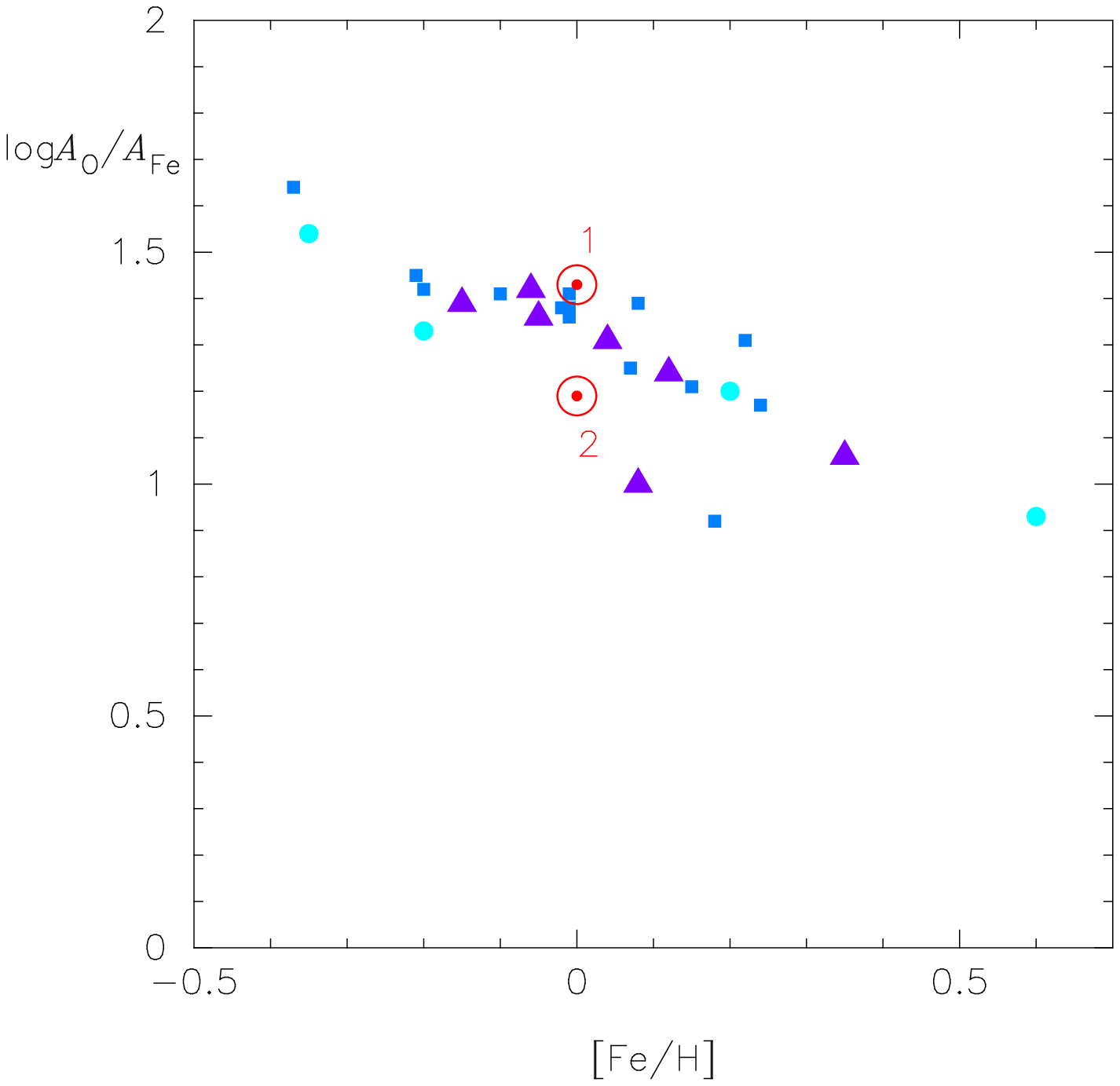}
   \end{center}
   \caption{The oxygen-to-iron ratios are plotted against [Fe/H] = 
log\,$A_{\rm Fe}^{*}$ - log\,$A_{\rm Fe}^{\odot}$ in M dwarfs and the Sun.
The oxygen-to-iron ratio of the classical oxygen abundance \citep{And89} 
follows the general trend shown by the M dwarfs ($\odot$ marked with 1), 
but that of the recent downward revised result \citep{Asp09} does not 
follow the general trend ($\odot$ marked with 2). Thus, if the recently
recommended low oxygen abundance is correct, the solar oxygen abundance 
is atypical for its metallicity. }
\label{figure16}
\end{figure}

\begin{figure}
   \begin{center}
       \FigureFile(80mm,65mm){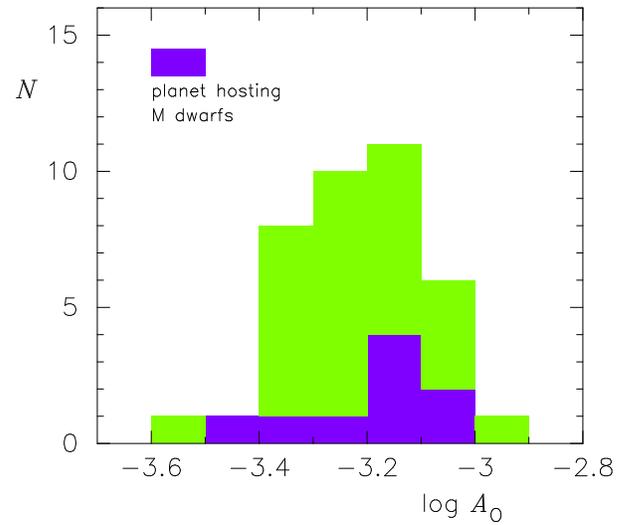}
   \end{center}
   \caption{Frequency distribution of M dwarfs against log\,$A_{\rm O}$.
Note that nine  M dwarfs are  hosting planet(s).    }
\label{figure17}
\end{figure}

\onecolumn

\begin{figure}
   \begin{center}
       \FigureFile(160mm,115mm){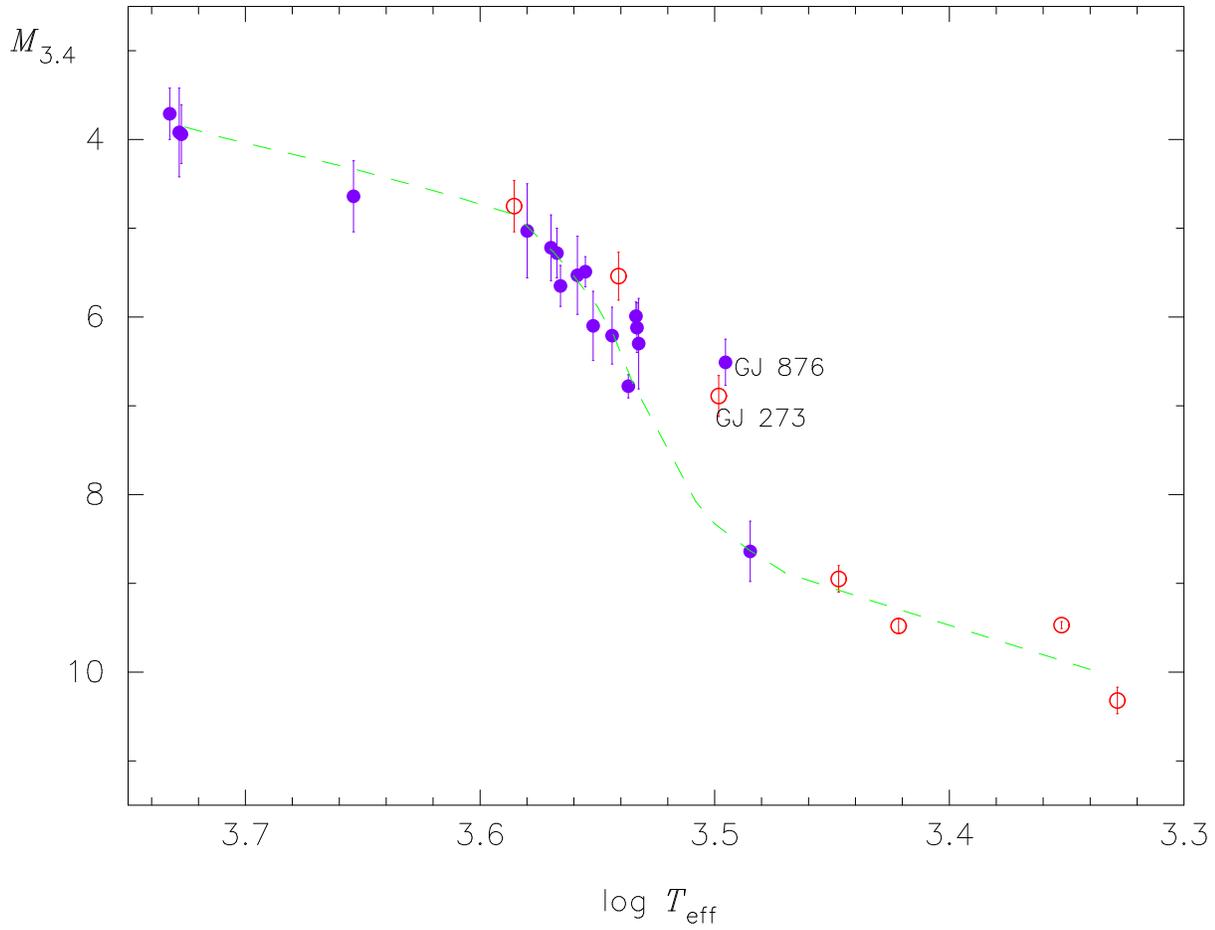}
   \end{center}
   \caption{The absolute magnitudes at 3.4\,$\mu$m, $M_{\rm 3.4}$, plotted 
against log\,$T_{\rm eff}$,  where $M_{\rm 3.4}$ values are based on the 
AllWISE Catalog (see Tables\,10 and 11) instead of the WISE All-Sky 
Catalog applied in Paper I. The probable errors of the AllWISE Catalog
are larger than those of the WISE All-Sky Catalog in general. 
The values of $T_{\rm eff}$  based on 
the interferometry  and infrared flux method are shown by the filled 
and open circles, respectively.  The mean $M_{\rm 3.4}$ -- log\,$T_{\rm eff}$
relation shown by the dashed line is exactly the same as that in Fig.\,1 
of Paper I. 
}\label{figure18}
\end{figure}

\newpage

\begin{table*}
\begin{center}
\caption{Fundamental Parameters and Model Photospheres for the Program Stars.}
\begingroup
\renewcommand{\arraystretch}{0.72}
\begin{tabular}{llccc}
\hline
obj.        & Sp. Type\footnotemark[$*$]  & $ T_{\rm eff}$\footnotemark[$*$] &
  log\,$g$\footnotemark[$*$] & model\footnotemark[$*$] \\
\hline
            &        &       &         &               \\
GJ\,15A     & dM2.5  &  3567 &   4.890  &  Ca3570c489  \\
GJ\,105B    & dM4.5  &  3360 &   4.954  &  Ca3360c495  \\
GJ\,166C    & dM4e   &  3337 &   4.972  &  Ca3340c497  \\
GJ\,176     & dM2.5e &  3616 &   4.804  &  Ca3620c480  \\
GJ\,179     & dM3.5e &  3476 &   4.877  &  Ca3480c488  \\
GJ\,205     & dM3    &  3801 &   4.710  &  Ca3800c471  \\
\end{tabular}
\endgroup
\end{center}
\end{table*}

\begin{table*}
\begin{center}
\caption{H$_2$O Lines in the Region A. }
\begingroup
\renewcommand{\arraystretch}{0.82}
\begin{tabular}{lccccccccccrr}
\hline
Ref.\footnotemark[$*$] & $ \lambda_{\rm obs} $\footnotemark[$\dagger$] & $ \lambda $\footnotemark[$\dagger$] & 
 $v_1^{'}v_2^{'}v_3^{'}$ & $v_1^{''}v_2^{''}v_3^{''}$ & 
$J^{'}$ & $K_a^{'}$ & $K_c^{'}$ & $J^{''}$ & $K_a^{''}$ & $K_c^{''}$ &  log\,$gf$ & L.E.P.\footnotemark[$\ddagger$]  \\   
\hline
    &          &           &         &         &     &     &     &       &     &     &            &            \\
A01 &  20298.6 & 20298.213 & 1\,0\,1 & 0\,1\,0 &  15 &   8 &   8 &    16 &   8 &   9 &     -5.195 &   5949.217 \\
    &          & 20298.428 & 0\,3\,2 & 0\,2\,1 &   9 &   3 &   7 &    10 &   3 &   8 &     -5.049 &   8358.617 \\
    &          & 20298.838 & 0\,1\,1 & 0\,0\,0 &  19 &  11 &   9 &    20 &  11 &  10 &     -5.074 &   6664.142 \\
A02 &  20306.4 & 20306.355 & 0\,1\,1 & 0\,0\,0 &  17 &   0 &  17 &    18 &   0 &  18 &     -5.284 &   3319.448 \\
    &          & 20306.359 & 0\,1\,1 & 0\,0\,0 &  17 &   1 &  17 &    18 &   1 &  18 &     -4.676 &   3319.448 \\
A03 &  20308.6 & 20308.271 & 0\,1\,1 & 0\,0\,0 &  16 &   6 &  10 &    17 &   6 &  11 &     -4.909 &   4291.907 \\
\end{tabular}
\endgroup
\end{center}
\end{table*}

\begin{landscape}

\begin{table*}

\begin{center}
 
\caption{Log\,$(W/\lambda)_{\rm obs}$\footnotemark[$*$] of H$_2$O Blends in 
the Region A of 38 M Dwarfs\footnotemark[$\dagger$].}
\label{tab:seventh}

\scalebox{0.95}[0.6]{   
\begin{tabular}{lccccccccccccccccc}
\hline
obj.      &   A01 &   A02  &   A03  &   A04  &   A05  &   A06  &   A07  &   A08  &   A09  &   A10 &  A11  & A12 &  A13 &  A14 & A15 & A16 & A17 \\
   \hline
          &       &        &        &        &        &        &         &       &        &        &        &        &        &        &        &        &  \\
GJ15A  &  --- & -4.957 & -4.803 & -4.671 &  --- & -5.018 &  --- & -5.177 &  --- & -5.046 & -4.994 &  --- & -5.018 &  --- &  --- & -4.822 & -4.847 \\
GJ105B & -4.757 & -4.685 & -4.522 & -4.405 &  --- & -4.815 & -4.711 & -4.854 & -4.779 & -4.739 &  --- &  --- & -4.772 &  --- &  --- & -4.587 & -4.529 \\
GJ166C & -4.760 & -4.709 & -4.521 & -4.375 &  --- & -4.740 & -4.595 & -4.772 &  --- &  --- &  --- &  --- &  --- &  --- &  --- &  --- & -4.430 \\
GJ176  &  --- & -4.981 & -4.837 & -4.707 &  --- & -5.070 &  --- & -5.233 & -5.163 & -5.034 &  --- &  --- & -4.987 &  --- &  --- & -4.797 & -4.793 \\
GJ179  &  --- & -4.837 & -4.689 & -4.545 &  --- & -4.958 & -4.856 & -5.061 &  --- & -4.906 &  --- & -4.667 & -4.928 & -4.673 &  --- & -4.781 & -4.674 \\
GJ205  &  --- & -4.986 &  --- & -5.026 &  --- &  --- &  --- &  --- &  --- &  --- &  --- &  --- &  --- &  --- & -5.260 & -5.120 & -5.173 \\
\end{tabular}
}    
\end{center}
\end{table*}
\end{landscape}


\begin{table*}
\begin{center}
\caption{ H$_2$O Lines in the Region B.}
\begingroup
\renewcommand{\arraystretch}{0.7}
\begin{tabular}{lccccccccccrr}
\hline
Ref.\footnotemark[$*$] & $ \lambda_{\rm obs} $\footnotemark[$\dagger$] & $ \lambda $\footnotemark[$\dagger$] & 
 $v_1^{'}v_2^{'}v_3^{'}$ & $v_1^{''}v_2^{''}v_3^{''}$ & 
$J^{'}$ & $K_a^{'}$ & $K_c^{'}$ & $J^{''}$ & $K_a^{''}$ & $K_c^{''}$ &  log\,$gf$ & L.E.P.\footnotemark[$\ddagger$]  \\   
\hline
    &          &           &         &         &     &     &     &       &     &     &            &            \\   
B01 &  22517.5 & 22517.619 & 0\,2\,0 & 0\,0\,0 &  17 &   8 &   9 &    16 &   5 &  12 &     -5.910 &   3639.538 \\
    &          & 22517.760 & 1\,0\,0 & 0\,0\,0 &  18 &   5 &  14 &    17 &   2 &  15 &     -5.907 &   3567.174 \\
B02 &  22528.7 & 22528.588 & 0\,1\,1 & 0\,1\,0 &  18 &   8 &  10 &    17 &   6 &  11 &     -5.283 &   5965.714 \\
    &          & 22528.871 & 1\,0\,0 & 0\,0\,0 &  20 &   7 &  14 &    19 &   4 &  15 &     -5.410 &   4851.821 \\
B03 &  22537.6 & 22537.447 & 1\,0\,0 & 0\,0\,0 &  19 &   6 &  14 &    18 &   3 &  15 &     -6.133 &   4201.253 \\
    &          & 22537.570 & 0\,0\,1 & 0\,0\,0 &  17 &   7 &  11 &    16 &   5 &  12 &     -5.765 &   3639.538 \\
\end{tabular}
\endgroup
\end{center}
\end{table*}

{\tiny

\renewcommand{\tabcolsep}{3pt}

\begin{landscape}

\begin{table*}

\begin{center}
 
\caption{ Log\,$(W/\lambda)_{\rm obs}$\footnotemark[$*$] of H$_2$O Blends 
 in the Region B of 20 M Dwarfs\footnotemark[$\dagger$].}
\label{tab:seventh}

\scalebox{0.75}{   
\begin{tabular}{lccccccccccccccccccccccccccc}
\hline
obj.      &   B01  &   B02  &   B03  &   B04  &   B05  &   B06  &   B07  &   B08  &   B09  &  B10 &   B11  & B12 & B13 &  B14 &  B15 & B16 &  B17  &
B18  &  B19  & B20  & B21 & B22 & B23 & B24 & B25 & B26 & B27 \\
   \hline
       &       &       &     &       &     &     &    &       &       &      &        &        &    &      &       &      &       &    
     &        &        &        &        &        &        &        &        &      \\
GJ105B      & -5.229 & -5.078 & -5.268 &  --- &  --- & -5.476 &  --- &  --- &  --- &  --- & -5.262 &  --- & -5.152 & -5.128 &  --- & -5.451 & -5.245 & -5.322 &  --- &  --- &  --- & -5.059 &  --- &  --- &  --- & -5.049 & -5.097 \\ 
GJ166C      & -5.191 & -5.060 & -5.351 &  --- &  --- &  --- & -4.884 &  --- & -5.172 & -5.216 & -5.274 &  --- & -5.140 & -5.133 &  --- & -5.244 & -5.081 & -5.138 &  --- & -4.981 &  --- & -5.105 &  --- & -4.910 & -5.322 & -5.017 & -5.038 \\
GJ179       & -5.439 & -5.321 & -5.470 & -5.430 &  --- & -5.592 & -5.194 &  --- &  --- &  --- & -5.416 &  --- & -5.522 & -5.429 &  --- & -5.518 & -5.186 & -5.612 &  --- &  --- &  --- &  --- & -5.376 & -5.087 &  --- & -5.111 & -5.218  \\
GJ231.1B & -5.421 & -5.307 & -5.432 & -5.322 &  --- &  --- & -5.189 &  --- & -5.433 &  --- & -5.290 &  --- &  --- & -5.102 &  --- & -5.432 & -5.302 & -5.495 &  --- &  --- &  --- &  --- &  --- & -5.365 & -5.563 & -5.212 & -5.367  \\
GJ273       & -5.526 & -5.249 & -5.320 & -5.259 &  --- &  --- & -5.023 &  --- & -5.256 &  --- & -5.320 &  --- &  --- & -5.269 & -5.341 &  --- & -5.209 & -5.437 &  --- & -5.239 & -5.412 & -5.452 &  --- &  --- & -5.363 & -5.151 & -5.190   \\
GJ324B   　 &  --- & -5.297 & -5.269 & -5.399 &  --- & -5.555 & -5.207 &  --- & -5.505 &  --- &  --- &  --- & -5.328 & -5.548 &  --- & -5.539 & -5.226 & -5.555 &  --- & -5.306 &  --- & -5.593 & -5.446 &  --- & -5.670 & -5.267 & -5.345  \\
\end{tabular}
}    
 
\end{center}
\end{table*}
\end{landscape} 

}   

\renewcommand{\tabcolsep}{4pt}

\begin{table*}
\begin{center}
\caption{Oxygen Abundances (log\,$A_{\rm O}^{A}$) from the H$_2$O Lines in the Region A. }
\begingroup
\renewcommand{\arraystretch}{0.68}
\begin{tabular}{lccccccrrrr}
\hline
obj. &  log\,$A_{\rm C}$\footnotemark[$*$] & $\Delta{\rm log}\,A_{\rm O}^{A}$\footnotemark[$\dagger$] 
& log\,$A_{\rm O}^{A} \pm {\rm p.e.} $ & $N_{A}$\footnotemark[$\ddagger$] &
$\chi^{2}_{-}$\footnotemark[$\S$] & $\chi^{2}_{0}$\footnotemark[$\|$] &  $\chi^{2}_{+}$\footnotemark[$\#$]  \\
\hline
            &                &          &                     &      &        &        &        \\
GJ\,15A     &    -3.601 &  -0.012 &  -3.313 $\pm$ 0.013 &  10  &  3.947 &  2.644 &  3.987 \\
GJ\,105B    &    -3.468 &  +0.026 &  -3.142 $\pm$ 0.033 &  12  &   7.605 &  7.554 &  9.413 \\
GJ\,166C    &    -3.366 &  +0.039 &  -3.027 $\pm$ 0.043 &   8  &  14.971 & 12.498 & 11.775 \\
GJ\,176     &    -3.352 &  -0.101 &  -3.153 $\pm$ 0.021 &  10  &   8.747 &  4.841 &  7.720 \\
GJ\,179     &    -3.427 &  -0.074 &  -3.201 $\pm$ 0.021 &  12  &   7.146 &  4.549 &  6.724 \\
GJ\,205     &    -3.116 &  -0.152 &  -2.968 $\pm$ 0.046 &   5  &   2.519 &  4.037 &  9.772 \\
\end{tabular}
\endgroup
\end{center}
\end{table*}

\begin{table*}
\begin{center}
\caption{Oxygen Abundances (log\,$A_{\rm O}^{B}$) from the H$_2$O Lines in the Region B. }
\begin{tabular}{lrcrcrr}
\hline
obj. & $ \Delta {\rm log}\,A_{\rm O}^{A}$\footnotemark[$*$] & log\,$A_{\rm O}^{A} 
\pm {\rm p.e.}$\footnotemark[$*$] &  $\Delta {\rm log}\,A_{\rm O}^{B}$\footnotemark[$\dagger$] & 
 log\,$A_{\rm O}^{B}\pm {\rm p.e.}$ & $N_{B}$\footnotemark[$\ddagger$] & diff\footnotemark[$\S$]~  \\ 
\hline
            &             &                     &            &                     &    &       \\ 
GJ\,105B    &    +0.026~~ &  -3.142 $\pm$ 0.033 &   +0.040~~ & -3.128 $\pm$ 0.050  & 13 &  +0.014 \\ 
GJ\,166C    &    +0.039~~ & -3.027 $\pm$ 0.043  &   +0.033~~ & -3.033 $\pm$ 0.074  & 18 &  -0.006 \\ 
GJ\,179     &    -0.074~~ & -3.201 $\pm$ 0.021  &   -0.044~~ & -3.171 $\pm$ 0.062  & 16 &  +0.030 \\ 
GJ\,231.1B  &    +0.037~~ & -3.235 $\pm$ 0.035  &   +0.028~~ & -3.244 $\pm$ 0.064  & 15 &  -0.009 \\ 
GJ\,273     &    -0.015~~ & -3.118 $\pm$ 0.023  &   -0.034~~ & -3.137 $\pm$ 0.047  & 17 &  -0.019 \\ 
GJ\,324B    &    -0.103~~ & -3.163 $\pm$ 0.041  &   -0.112~~ & -3.172 $\pm$ 0.040  & 17 &  -0.009 \\
\end{tabular}
\end{center}
\end{table*}

\twocolumn
    
\begin{table}
\begin{center}
\caption{Carbon and Oxygen Abundances in M Dwarfs.  }
\begingroup
\renewcommand{\arraystretch}{0.65}
\begin{tabular}{lclc}
\hline
obj. &  log\,$A_{\rm C} \pm {\rm p.e.}$\footnotemark[$*$] &   
log\,$A_{\rm O} \pm {\rm p.e.}$\footnotemark[$\dagger$] & 
log\,${A_{\rm O}/A_{\rm C}}$   \\
\hline
             &                    &                    &               \\
GJ~15A        &   -3.60 $\pm$ 0.11 &   -3.31 $\pm$ 0.01A &    0.29        \\
GJ~105B       &   -3.47 $\pm$ 0.06 &   -3.14 $\pm$ 0.04 &    0.33        \\
GJ~166C       &   -3.37 $\pm$ 0.14 &   -3.03 $\pm$ 0.06 &    0.34        \\
GJ~176\footnotemark[$\ddagger$]        &   -3.35 $\pm$ 0.07 &   -3.15 $\pm$ 0.02A &    0.20        \\
GJ~179\footnotemark[$\ddagger$]        &   -3.43 $\pm$ 0.11 &   -3.18 $\pm$ 0.04 &    0.24        \\
GJ~205        &   -3.12 $\pm$ 0.08 &   -2.97 $\pm$ 0.05A &    0.15        \\
\end{tabular}
\endgroup
\end{center}
\end{table}

\begin{table}
\begin{center}
\caption{O/Fe Ratios in M Dwarfs.  }
\begingroup
\renewcommand{\arraystretch}{1.35}
\begin{tabular}{llllc}
\hline
obj. &  [Fe/H]\footnotemark[$*$]  & log\,$A_{\rm Fe}$ &
log\,$A_{\rm O}$\footnotemark[$\dagger$] &
log\,${A_{\rm O}/A_{\rm Fe}}$   \\
\hline
GJ\,15A  &   -0.35\footnotemark[$\ddagger$]  &  -4.85 & -3.31 &  1.54 \\
GJ\,205  &   +0.6\footnotemark[$\S$]   &  -3.90 & -2.97 &  0.93 \\
GJ\,229  &   +0.2\footnotemark[$\ddagger$]   &  -4.30 & -3.10 &  1.20 \\
GJ\,411  &   -0.2\footnotemark[$\S$]   &  -4.70 & -3.37 &  1.33 \\
GJ\,105B  &  -0.06\footnotemark[$\|$]  &  -4.56 & -3.14 &  1.42 \\ 
GJ\,176   &  +0.04\footnotemark[$\|$]  &  -4.46 & -3.15 &  1.31 \\
GJ\,250B  &  -0.05\footnotemark[$\|$]  &  -4.55 & -3.19 &  1.36 \\
\end{tabular}
\endgroup
\end{center}
\end{table}

\onecolumn

\begin{table*}
\begin{center}
\caption{Fundamental Parameters Based on $T_{\rm eff}$ by the Interferometry.}
\begin{tabular}{llcccccc}
   \hline
obj. & sp.ty.\footnotemark[$*$] & $p$(msec)\footnotemark[$\dagger$] &
$F_{3.4}$(mag)\footnotemark[$\ddagger$]  & $M_{3.4}$(mag)\footnotemark[$\S$] &
$F_{3.4}$(mag)\footnotemark[$\|$]  & $M_{3.4}$(mag)\footnotemark[$\S$]
   & $T_{\rm eff}$    \\
   \hline
       &       &                   &                   &             &        &     &  \\
GJ~15A & dM2.5 & 278.76 $\pm$~0.77 & 3.853 $\pm$~0.099 & 6.08 $\pm$~0.10 &
3.872 $\pm$~0.381 & 6.10 $\pm$~0.39 & 3563 $\pm$~11  \\
GJ~53A & G5Vp & 132.38 $\pm$~0.82 & 3.326 $\pm$~0.141 & 3.94 $\pm$~0.15 &
 3.311 $\pm$~0.487 & 3.92 $\pm$~0.50 & 5348 $\pm$~26  \\
GJ~ 75 & K0V &  99.33 $\pm$~0.53 &  3.753$\pm$~0.091 & 3.74 $\pm$~0.10 &
 3.725 $\pm$~0.278 & 3.71 $\pm$~0.29  & 5398 $\pm$~75 \\
GJ~176\footnotemark[$\#$] & dM2.5e & 107.83 $\pm$~2.85 & 5.434 $\pm$~0.066 & 
5.60 $\pm$~0.12 & 5.486 $\pm$~0.177 & 5.65 $\pm$~0.23 & 3679 $\pm$~77     \\
GJ~205 & dM3 & 176.77 $\pm$~1.18 & 3.743 $\pm$~0.120 & 4.98 $\pm$~0.13 &
 3.790 $\pm$~0.519 & 5.03 $\pm$~0.53 & 3801 $\pm$~ 9  \\
GJ~412A & dM2 & 206.27 $\pm$~1.00 & 4.638 $\pm$~0.085 & 6.21 $\pm$~0.10 &
 4.641 $\pm$~0.312 & 6.21 $\pm$~0.32  & 3497 $\pm$~39  \\
\end{tabular}
\end{center}
\end{table*}


\begin{table*}
\begin{center}
\caption{Fundamental Parameters Based on $T_{\rm eff}$ by the Infrared Flux 
Method.}
\begin{tabular}{llcccccc}
   \hline
obj. & sp.ty.\footnotemark[$*$] & $p$(msec)\footnotemark[$\dagger$] &
$F_{3.4}$(mag)\footnotemark[$\ddagger$]  & $M_{3.4}$(mag)\footnotemark[$\S$]  &
$F_{3.4}$(mag)\footnotemark[$\|$]  & $M_{3.4}$(mag)\footnotemark[$\S$]
 & $T_{\rm eff}$\footnotemark[$\#$]    \\
   \hline
       &       &             &                   &                 &      &        &          \\
GJ~273-L\footnotemark[$**$] & dM4 & 262.98 $\pm$~1.39 & 4.723 $\pm$~0.074 & 6.82 $\pm$~0.09 & 
 4.787 $\pm$~0.223 & 6.89 $\pm$~0.23 & 3150 $\pm$~95  \\
GJ~406 & dM6.5e & 419.10 $\pm$~2.10 & 5.807 $\pm$~0.055 & 8.92 $\pm$~0.07 & 
 5.836 $\pm$~0.138 & 8.95 $\pm$~0.15 & 2800 $\pm$~85  \\
GJ~644C & M7V & 148.92 $\pm$~4.00 & 8.588 $\pm$~0.023 & 9.45 $\pm$~0.08 &
  8.619 $\pm$~0.023 & 9.48 $\pm$~0.08 & 2640 $\pm$~80 \\
GJ~752A & M3V & 170.36 $\pm$~1.00 & 4.466 $\pm$~0.078 & 5.62 $\pm$~0.09 &
 4.379 $\pm$~0.262 & 5.54 $\pm$~0.27 & 3475$ \pm$~105 \\
GJ~752B & M8V & 170.36 $\pm$~1.00 & 8.465 $\pm$~0.023 & 9.62 $\pm$~0.04 &
 8.317 $\pm$~0.023 & 9.47 $\pm$~0.04  & 2250 $\pm$~70 \\
GJ~884 & dM0.5 & 121.69 $\pm$~0.69 & 4.424 $\pm$~0.087 & 4.85 $\pm$~0.10 &
 4.324 $\pm$~0.273 & 4.75 $\pm$~0.29  & 3850 $\pm$~115  \\
\end{tabular}
\end{center}
\end{table*}


\vspace{2mm}

\begin{thebibliography}{}

\bibitem[Allende Prieto et al.(2002)]{All02}
  Allende Prieto, C., Lambert, D. L., \& Asplund, M.\ 2002, \apj, 573, L137
\bibitem[Anders \& Grevesse(1989)]{And89}
  Anders, E., \& Grevesse, N.\ 1989, Geochim. Cosmochim. Acta, 53, 197
\bibitem[Asplund et al. (2009)]{Asp09}
  Asplund, M., Grevesse, N., Sauval, A. J., \& Scott, P.\ 2009, \araa, 47, 481
\bibitem[Auman(1967)]{Aum67}
  Auman, J. R.\ 1967, \apjs, 14, 171
\bibitem[Auman(1969)]{Aum69}
  Auman, J. R.\ 1969, in Low-Luminosity Stars ed. S. S. Kumar (New York: 
  Gordon and Breach Sci. Publ.), 483
\bibitem[Barber et al.(2006)]{Bar06}
  Barber, R. J., Tennyson, J., Harris, G. J., \& Tolchenov, R. N.\ 2006,
    \mnras, 368, 1087
\bibitem[Bean et al.(2006)]{Bea06}
  Bean, J. L., Sneden, C., Hauschildt, P. H., Johns-Krull, C. M., \&
  Benedit, G. F.\ 2006, \apj,  652, 1604
\bibitem[Bernath (2002)]{Ber02}
  Bernath, P. F.\ 2002, Phys. Chem. Chem. Phys., 4, 1501
\bibitem[Berriman \& Reid(1987)]{Ber87}
  Berriman, G., \& Reid, N.\ 1987, \mnras, 227, 315
\bibitem[Bi\'emont et al.(1991)]{Bie91}
  Bi\'emont, E., Baudoux, M., Kurucz, R. L., Ansbacher, W., \& 
 Pinnington, E. H.\ 1991, \aap, 249, 539
\bibitem[Boyajian et al.(2012)]{Boy12}
  Boyajian, T. S., et al.\ 2012, \apj, 757, 112
\bibitem[Connes(1970)]{Con70}
  Connes, P.\ 1970, \araa, 8, 209
\bibitem[Cox(1999)]{Cox99}
  Cox, A. N. (ed.)\ 1999, Allen's Astrophysical Quantities, Fourth Edition
  (Berlin: Springer-Verlag), 145
\bibitem[Decin et al.(2003)]{Dec03}
  Decin, L., Vandenbussche, B., Waelkens, C., Decin, G., Eriksson, K.,
  Gustafsson, B., Plez, B., \& Sauval, A. J.\ 2003, \aap, 400, 709 
\bibitem[Edvardsson et al.(1993)]{Edv93}
   Edvardsson, B., Andersen, J., Gustafsson, B., Lambert, D. L., Nissen,
   P. E., \& Tomkin, J.\ 1993, \aap, 275, 101
\bibitem[Faure et al.(2013)]{Fau13}
  Faure, A., Wiesenfeld, L., Drouin, B. J., \& Tennyson, J.\ 2013, JQSRT,
  116, 79
\bibitem[Fischer \&  Valenti(2005)]{Fis05}
  Fischer, D. A., \& Valenti, J. A.\ 2005, \apj, 622, 1102
\bibitem[Gamache et al.(1996)]{Gam96}
  Gamache, R. R., Lynch, R., \& Brown, L. R.\ 1996, JQSRT, 56, 471
\bibitem[Gliese \& Jahreiss(1991)]{Gli91}
  Gliese, W., \& Jahreiss, H.\ 1991, Preliminary Version of the Third
  Catalogue of Nearby Stars (Greenbelt: NASA/Astronomical Data Center)
\bibitem[Grevesse \& Sauval(1999)]{Gre99}
  Grevesse, N., \& Sauval, A. J.\ 1999, \aap, 347, 348
\bibitem[Gustafsson et al.(1999)]{Gus99}
   Gustafsson, B., Karlsson, T., Olsson, E., Edvardsson, B., \& Ryde, N.\ 
  1999, \aap, 342, 426
\bibitem[Hattori et al.(2014)]{Hat14}
  Hattori, K., Yoshii, Y., Beers, T., Carollo, D., \& Lee, Y. S.\ 2014, 
  \apj, 784, 153
\bibitem[Herzberg(1945)]{Her45}
 Herzberg, G.\ 1945, Molecular Spectra and Molecular Structure II. Infrared
 and Raman Spectra of Polyatomic Moleules (New York: D. Van Nostrand Co. Inc.),
 207
\bibitem[Hinkle \& Barnes (1979)]{Hin79}
   Hinkle, K. H., \& Barnes, T. G.\ 1979, \apj, 227, 923
\bibitem[Holweger et al.(1990)]{Hol90}
  Holweger, H., Heise, C., \& Kock, M.\ 1990, \aap, 232, 510
\bibitem[Johnson \& Apps (2009)]{Joh09}
  Johnson, J. A., \& Apps, K.\ 2009, \apj, 699, 933
\bibitem[Jones et al.(1994)]{Jon94}
  Jones, H. R. A., Longmore, A. J., Jameson, R. F., \& Mountain, C. M. 1994,
  \mnras, 267, 413 
\bibitem[Joy \& Abt (1974)]{Joy74}
  Joy, A. H., \& Abt, H. A.\ 1974, \apjs, 28, 1
\bibitem[Kessler et al.(1996)]{Kes96}
   Kessler, M. F., et al.\ 1996, \aap, 315, L27
\bibitem[King et al.(1943)]{Kin43}
  King, G. W., Hainer, R. M., \& Cross, P. C.\ 1943, \jcp, 11, 27
\bibitem[Kobayashi et al.(2000)]{Kob00}
  Kobayashi, N., et al.\ 2000, Proc. SPIE, 4008, 1056
\bibitem[Kuiper(1962)]{Kui62}
  Kuiper, G. P.\ 1962, Comm. Lunar Planet. Lab., No.23, 179
\bibitem[Lebzelter et al.(2012)]{Leb12}
 Lebzelter, T., et al.\ 2012, \aap, 547, A108
\bibitem[Mohler(1955)]{Moh55}
  Mohler, O. C.\ 1955, A Table of Solar Spectrum Wave Lengths 11984A to
  25578A (Ann Arbor: The University of Michigan Press)
\bibitem[Montarg\`es et al.(2014)]{Mon14}
   Montarg\`es, M., Kervella, P., Perrin, G., Ohnaka, K., Chiavassa, A., 
   Ridgway, S. T., \& Lacour, S.\ 2014, \aap, 572, A17
\bibitem[Mould(1978)]{Mou78}
 Mould, J. R.\ 1978, \apj, 226, 923
\bibitem[Neves et al.(2013)]{Nev13}
  Neves, V., Bonfils, X., Santos, N. C., Delfosse, X., Forveille, T.,
Allard, F., \&  Udry, S.\ 2013, \aap, 551, A36
\bibitem[Nomoto et al.(2013)]{Nom13}
   Nomoto, K., Kobayashi, C., \& Tominaga, N.\ 2013, \araa, 51, 457
\bibitem[Ohnaka(2004)]{Ohn04}
   Ohnaka, K.\ 2004, \aap, 421, 1149
\bibitem[Ohnaka et al.(2011)]{Ohn11}
   Ohnaka, K., et al.\ 2011, \aap, 529, A163
\bibitem[\"Onehag et al.(2012)]{One12}
 \"Onehag, A., Heiter, U., Gustafsson, B., Piskunov, N., Plez, B., \&
 Reiners, A.\ 2012, \aap, 542, A33
\bibitem[Polyansky et al.(1997)]{Pol97}
  Polyansky, O. L., Zobov, N. F., Viti, S., Tennyson, J., Bernath, P. F.,  
  \& Wallace, L.\ 1997, \apj, 489, L205
\bibitem[Ridgway \& Brault (1984)]{Rid84}
  Ridgway, S. T., \& Brault, J. W. 1984, \araa, 22, 29
\bibitem[Rothman et al.(2010)]{Rot10}
  Rothman, L. S., et al.\ 2010, JQSRT, 111, 2139
\bibitem[Russell(1934)]{Rus34}
  Russell, H. N.\ 1934, \apj, 79, 317 
\bibitem[Ryde et al.(2002)]{Ryd02}
  Ryde, N., Lambert, D. L., Richter, M. J., \& Lacy, J. H.\ 2002, 
  \apj, 580, 447  
\bibitem[Steyert et al.(2004)]{Ste04}
  Steyert, D. W., Wang, W. F., Sirota, J. M., Donahue, N. M.,  \& 
  Reuter, D. C.\ 2004, JQSRT, 83, 183
\bibitem[Struve \& Elvey(1934)]{Str34}
  Struve, O., \& Elvey, C. T.\ 1934, \apj, 79, 409
\bibitem[Tereszchuk et al.(2002)]{Ter02}
  Tereszchuk, K., Bernath, P. F., Zobov, N. F., Shirin, S. V., Polyansky, 
  O. L., Libeskind, N. I., Tennyson, J., \& Wallace, L.\ 2002, \apj, 577, 496
\bibitem[Tinney et al.(1993)]{Tin93}
   Tinney, C. G., Mould, J. R., \& Reid, I. N.\ 1993, \aj, 105, 1045 
\bibitem[Tsuji(1966)]{Tsu66}
  Tsuji, T.\ 1966, \pasj, 18, 127
\bibitem[Tsuji(1969)]{Tsu69}
  Tsuji, T.\ 1969, in Low-Luminosity Stars ed. S. S. Kumar (New York:
  Gordon and Breach Sci. Publ.), 457
\bibitem[Tsuji(2000)]{Tsu00}
   Tsuji, T.\ 2000, \apj, 538, 801
\bibitem[Tsuji(2001)]{Tsu01}
   Tsuji, T.\ 2001, \aap, 376, L1
\bibitem[Tsuji(2002)]{Tsu02}
   Tsuji, T.\ 2002, \apj, 575, 264
\bibitem[Tsuji(2005)]{Tsu05}
   Tsuji, T.\ 2005, \apj, 621, 1033
\bibitem[Tsuji(2008)]{Tsu08}
   Tsuji, T.\ 2008, \aap, 489, 1271
\bibitem[Tsuji \& Nakajima(2014)]{Tsu14}
  Tsuji, T., \& Nakajima, T.\ 2014, \pasj, 66, 98 (Paper I)
\bibitem[Tsuji et al.(1996)]{Tsu96}
 Tsuji, T., Ohnaka, K., \& Aoki, W.\ 1996, \aap, 305, L1
\bibitem[Uns\"old(1955)]{Uns55}
 Uns\"old, A.\ 1955, Physik der Sternatmosph\"aren mit Besonderer 
 Ber\"ucksichtigung der Sonne, 2ten Auf. (Berlin: Springer-Verlag) 
\bibitem[van Leeuwen(2007)]{Lee07}
 van Leeuwen, F.\ 2007, \aap, 474, 653
\bibitem[von Braun et al.(2014)]{Bra14}
  von Braun, K., et al.\ 2014, \mnras, 438, 2413
\bibitem[Wallace et al.(1995)]{Wal95}
  Wallace, L., Bernath, P., Livingston, W., Hinkle, K., Busler, J., Guo, B.,
 \& Zhang, K.\ 1995, Science, 268, 1155
\bibitem[Wallace \& Livingston (1992)]{Wal92}
  Wallace, L., \& Livingston, W.\ 1992, N. S. O. Technical Report No.92-001,
  An Atlas of a Dark Sunspot Umbral Spectrum in the Infrared from 1970 to 
  8640 cm$^{-1}$ (Tucson: NOAO)
\bibitem[Woolf et al.(1964)]{Woo64}
  Woolf, N. J., Schwarzschild, M., \& Rose, W. K.\ 1964, \apj, 140, 833 
\bibitem[Wright et al.(2010)]{Wri10}
  Wright, E. L., et al.\ 2010, \aj, 140, 1868
\bibitem[Yoshii(1981)]{Yos81}
  Yoshii, Y.\ 1981, \aap, 97, 280 
\bibitem[Zobov et al.(2000)]{Zob00}
  Zobov, N. F., et al.\ 2000, \apj, 530, 994 
\bibitem[Zobov et al.(2008)]{Zob08}
  Zobov, N. F., et al.\ 2008, \mnras, 387, 1093 


\end{thebibliography}
\end{document}